\documentclass[aps, twocolumn,superscriptaddress,showkeys,amsmath,amssymb,floatfix,nofootinbib]{revtex4}%preprint <-> twocolumn  %,showpacs
\usepackage{booktabs,graphicx,mathrsfs,verbatim}
\usepackage[colorlinks=true,citecolor=blue,filecolor=blue,linkcolor=blue,urlcolor=blue,pdftex]{hyperref}
\usepackage[usenames,dvipsnames]{color}
\usepackage{ulem}
\usepackage{fancyhdr}
\usepackage{datetime}
\usepackage{multirow}

\newcommand{\1}[1]{\, \mathrm{#1}} % unit(y ;-)
\newcommand{\n}[1]{\mathrm{#1}} % normal (roman) text in math mode

\newcommand{\arxiv}[1]{\href{http://arxiv.org/abs/#1}{\texttt{arXiv:#1}}}

\newcommand{\dRdE}{\ensuremath{\frac{dR}{dE_{\n{nr}}}}}

\newcommand{\Poi}{\ensuremath{\textrm{Poi}}}

\newcommand{\Binom}{\textrm{Binom}}

\newcommand{\probpmti}{\ensuremath{p_{\mathrm{pmt},i}}}
\newcommand{\probpmt}{\ensuremath{p_{\mathrm{pmt}}}}
\newcommand{\Nph}{\ensuremath{N_{\gamma}}}
\newcommand{\nph}{\ensuremath{n_{\gamma}}}

\newcommand{\Nei}{\ensuremath{N_{pe,i}}}

\newcommand{\Wph}{\ensuremath{W_{\gamma}}}
\newcommand{\Leff}{\ensuremath{{\cal L}_{\mathrm{eff}}}}
\newcommand{\Efield}{\ensuremath{\mathcal{E}}}
\renewcommand{\|}{\ensuremath{\,|\,}}

\newcommand{\assergi}{\affiliation{INFN, Laboratori Nazionali del Gran Sasso, Assergi, Italy}}
\newcommand{\bern}{\affiliation{Albert Einstein Center for Fundamental Physics, University of Bern, Bern, Switzerland}}
\newcommand{\bologna}{\affiliation{Department of Physics and Astronomy, University of Bologna and INFN-Bologna, Bologna, Italy}}
\newcommand{\columbia}{\affiliation{Physics Department, Columbia University, New York, NY, USA}}
\newcommand{\coimbra}{\affiliation{Department of Physics, University of Coimbra, Coimbra, Portugal}}
\newcommand{\heidelberg}{\affiliation{Max-Planck-Institut f\"ur Kernphysik, Heidelberg, Germany}}
\newcommand{\houston}{\affiliation{Department of Physics and Astronomy, Rice University, Houston, TX, USA}}
\newcommand{\losangeles}{\affiliation{Physics \& Astronomy Department, University of California, Los Angeles, CA, USA}}
\newcommand{\mainz}{\affiliation{Institut f\"ur Physik \& PRISMA Exzellenzcluster, Johannes Gutenberg Universit\"at Mainz, Mainz, Germany}}
\newcommand{\munster}{\affiliation{Institut f\"ur Kernphysik, Wilhelms-Universit\"at M\"unster, M\"unster, Germany}}
\newcommand{\nikhef}{\affiliation{Nikhef and the University of Amsterdam, Science Park, Amsterdam, Netherlands}}
\newcommand{\purdue}{\affiliation{Department of Physics, Purdue University, West Lafayette, IN, USA}}
\newcommand{\shanghai}{\affiliation{Department of Physics and Astronomy, Shanghai Jiao Tong University, Shanghai, China}}
\newcommand{\subatech}{\affiliation{SUBATECH, Ecole des Mines de Nantes, CNRS/In2p3, Universit\'e de Nantes, Nantes, France}}
\newcommand{\torino}{\affiliation{INFN-Torino and Osservatorio Astrofisico di Torino, Torino, Italy}}
\newcommand{\weizmann}{\affiliation{Department of Particle Physics and Astrophysics, Weizmann Institute of Science, Rehovot, Israel}}
\newcommand{\zurich}{\affiliation{Physics Institute, University of Z\"{u}rich,  Z\"{u}rich, Switzerland}}

\begin{document}
\title{Analysis of the  XENON100  Dark Matter Search Data}
%\date{Version 10.10 -- \today}

\author{E.~Aprile}\columbia
\author{M.~Alfonsi}\nikhef 
\author{K.~Arisaka}\losangeles
\author{F.~Arneodo}\assergi
\author{C.~Balan}\coimbra
\author{L.~Baudis}\zurich
\author{A.~Behrens}\zurich
\author{P.~Beltrame}\losangeles\weizmann
\author{K.~Bokeloh}\munster
\author{E.~Brown}\munster
\author{G.~Bruno}\assergi
\author{R.~Budnik}\columbia 
\author{J.~M.~R.~Cardoso}\coimbra
\author{W.-T.~Chen}\subatech
\author{B.~Choi}\columbia
\author{D.~B.~Cline}\losangeles
\author{H.~Contreras}\columbia
\author{J.~P.~Cussonneau}\subatech
\author{M.~P.~Decowski}\nikhef
\author{E.~Duchovni}\weizmann
\author{S.~Fattori}\mainz
\author{A.~D.~Ferella}\zurich\assergi
\author{W.~Fulgione}\torino
\author{F.~Gao}\shanghai
\author{M.~Garbini}\bologna
\author{K.-L.~Giboni}\columbia
\author{L.~W.~Goetzke}\columbia
\author{C.~Grignon}\mainz
\author{E.~Gross}\weizmann
\author{W.~Hampel}\heidelberg
\author{A.~Kish}\zurich
\author{J.~Lamblin}\subatech
\author{H.~Landsman}\weizmann
\author{R.~F.~Lang}\purdue\columbia
\author{M.~Le~Calloch}\subatech
\author{C.~Levy}\munster
\author{K.~E.~Lim}\columbia
\author{Q.~Lin}\shanghai
\author{S.~Lindemann}\heidelberg
\author{M.~Lindner}\heidelberg
\author{J.~A.~M.~Lopes}\coimbra
\author{K.~Lung}\losangeles
\author{T.~Marrod\'an~Undagoitia} \email{marrodan@mpi-hd.mpg.de} \zurich\heidelberg%[Email:]
\author{F.~V.~Massoli}\bologna
\author{Y.~Mei}\houston\mainz
\author{A.~J.~Melgarejo~Fernandez}\columbia
\author{Y.~Meng}\losangeles
\author{A.~Molinario}\torino
\author{E.~Nativ}\weizmann
\author{K.~Ni}\shanghai
\author{U.~Oberlack}\mainz\houston
\author{S.~E.~A.~Orrigo}\coimbra
\author{E.~Pantic} \email{pantic@physics.ucla.edu} \losangeles
\author{R.~Persiani}\bologna
\author{G.~Plante}\columbia
\author{N.~Priel}\weizmann
\author{A.~Rizzo}\columbia
\author{S.~Rosendahl}\munster
\author{J.~M.~F.~dos Santos}\coimbra
\author{G.~Sartorelli}\bologna
\author{J.~Schreiner}\heidelberg
\author{M.~Schumann}\zurich\bern
\author{L.~Scotto~Lavina}\subatech
\author{P.~R.~Scovell}\losangeles
\author{M.~Selvi}\bologna
\author{P.~Shagin}\houston
\author{H.~Simgen}\heidelberg
\author{A.~Teymourian}\losangeles
\author{D.~Thers}\subatech
\author{O.~Vitells}\weizmann
\author{H.~Wang}\losangeles
\author{M.~Weber}\heidelberg
\author{C.~Weinheimer}\munster

\collaboration{The XENON100 Collaboration}\noaffiliation

\begin{abstract}

The XENON100 experiment, situated in the Laboratori Nazionali del Gran Sasso,  aims at the direct detection of dark matter in the form of weakly interacting massive particles (WIMPs),  based on their interactions with xenon nuclei in an ultra low background dual-phase time projection chamber. This paper describes the general methods developed for the analysis of the XENON100 data. These methods have been used in the 100.9 and 224.6\,live days science runs from which results on spin-independent elastic, spin-dependent elastic and inelastic WIMP-nucleon cross-sections have already been reported.

\end{abstract}



\maketitle

%\linenumbers

%%%%%%%%%%%%%%%%%%%%%%%%%%%%%%%%%%%%%%%%%%%%%%%%%%
\section{Introduction}
\label{chap::Introduction}
%%%%%%%%%%%%%%%%%%%%%%%%%%%%%%%%%%%%%%%%%%%%%%%%%%

The XENON100 experiment aims to directly detect cold, non-baryonic dark matter which accounts for the majority of the matter in the Universe\,\cite{Jarosik:2010iu} according to plentiful astronomical and cosmological evidence. Among the most promising  dark matter candidates are weakly interacting massive particles (WIMPs), arising naturally in several models beyond the Standard Model of particle physics\,\cite{Jungman:1995df}. If  WIMPs are the dark matter particles, they could be directly detected via scattering off nuclei\,\cite{Goodman:1984dc}. In this paper, the complete analysis of the low-energy XENON100 data is presented focusing on {two long science runs}. {Using 100.9\,live days of data, the} XENON100 collaboration reported exclusion limits on  spin-independent elastic\,\cite{Aprile:2011hi} and inelastic\,\cite{Aprile:2011ts}  WIMP-nucleon scattering. {A second science run, with 224.9\,live days of data, was used to report exclusion limits on spin-independent \cite{Aprile:2012kx} and spin-dependent \cite{Aprile:2013doa} elastic WIMP-nucleon scattering.}

Following this introduction, Section II {describes} the XENON100 experiment and details the measurement process. Section III  describes the data set, reconstruction procedures and selection cuts used in the analysis. This is followed by estimates of  nuclear and electronic recoil backgrounds with control samples from  Monte Carlo simulations, calibration data and science data outside of the signal region.  Finally, a description of the observed event population and the comparison  with the expected background is presented.

%%%%%%%%%%%%%%%%%%%%%%%%%%%%%%%%%%%%%%%%%%%%%%%%%%
\section{The XENON100 Experiment}
\label{chap::XENON100}
%%%%%%%%%%%%%%%%%%%%%%%%%%%%%%%%%%%%%%%%%%%%%%%%%%

\subsection{Instrument description} % =================================

The XENON100 detector is filled with a total of 161\,kg of ultra pure liquid xenon (LXe) divided in two concentric cylindrical volumes. The inner target volume is a two-phase (liquid/gas) time projection chamber (TPC) of  30.5\,cm  height and 15.3\,cm radius containing a xenon mass of 62\,kg. It is optically separated from the surrounding LXe veto, closed on the bottom by a cathode mesh and on the top by a gate and an anode mesh, which provide the homogeneous electric fields required for the operation of the TPC. A technique similar to the one of a diving bell was chosen to keep the liquid in the TPC at a precise level between the gate and the anode meshes and to raise the liquid level in the veto (outside of the bell) above the TPC. This enables full enclosure of the TPC itself in an active LXe shield of $\sim$4\,cm thickness  for an efficient  suppression of external radioactive backgrounds. The XENON100 instrument and its subcomponents are explained in detail in~\cite{Aprile:2011dd} and only a short summary is given here.

A particle interaction in the LXe target creates both excited and ionized Xe-atoms, which combine with the surrounding atoms  to form excimers\,\cite{Schwentner85}. De-excitation of these excimers leads to a prompt xenon scintillation signal (S1), which is recorded by photomultiplier tubes (PMTs) placed below the target in the LXe and above in the gas phase. Due to the presence of an electric drift field of 530\,V/cm, a large fraction of the ionization electrons is drifted away from the interaction site in the TPC with a drift velocity $v_d \simeq$ 1.73\,mm/$\mu$s. Electrons which escape recombination and are not trapped by impurities are extracted from the liquid into the gas phase by a strong extraction field of $\sim$12\,kV/cm, and a light signal (S2) is generated by proportional scintillation in the gas\,\cite{Dolgoshein}. The S2 signal is detected by the same PMTs but is delayed by the drift time $t_d$, which is the time it takes the electrons  to drift from the interaction site to the liquid/gas interface.  3-dimensional event vertex reconstruction is achieved using $t_d$ and $v_d$ to reconstruct the $z$ position ($z=v_{d}t_d$) and the hit pattern on the PMTs in the gas phase to reconstruct  the ($x, y$) position. The ratio S2/S1 is different for electronic recoil events from interactions with the atomic electrons (from $\gamma$ and $\beta$ backgrounds), and for interactions with the nucleus itself (nuclear recoils from WIMPs or neutrons), and is used to discriminate the signal against background.

The PMTs are 1\,inch\,$\times$\,1\,inch Hamamatsu R8520, selected for high quantum efficiency (up to 32\%) and very low intrinsic radioactivity. 80~tubes are immersed in the LXe below the TPC to ensure high light collection. 98~PMTs are placed in the gas phase above the target, arranged in concentric rings with the outmost ring extending beyond the TPC edge for improved ($x,y$) position reconstruction.  The LXe layer surrounding the target is instrumented with 64 additional PMTs, observing the volumes above, below and on the sides of the TPC, and operating as an active LXe veto. 

All materials used for the construction of the detector were selected for low intrinsic radioactivity\,\cite{Aprile:2011ru}, in order to minimize the electronic recoil background and neutron background from $(\alpha,n)$ and spontaneous fission reactions. To further suppress external background, the TPC and the active veto are installed inside a passive shield. From the inside to the outside this consists  of 5\,cm of oxygen free high conductivity copper, 20\,cm of polyethylene, 5\,cm of lead with a low $^{210}$Pb concentration, and 15\,cm of standard lead. Three sides and the top of the shield have 20-cm-thick water or polyethylene shielding to further lower the neutron background, and the entire shield installation sits on a 20\,cm slab of polyethylene.  The detector is installed underground in the Laboratori Nazionali del Gran Sasso (LNGS) in Italy, at an average depth of 3\,600\,m water equivalent, which effectively reduces the muon flux by a factor of $10^6$ compared to the surface flux\,\cite{Aglietta2005}.

\subsection{Measurement process and event rates}\label{MeasurementProcess}  % --- --- --- ---

This section describes how light and charge are measured in XENON100.
Upper case latin letters in S1 and S2 are used for the actual measured quantities while lower case letters in  s1 and s2 denote expectation values. $P$ is a discrete probability and $p$  is a probability density function (pdf). 

\subsubsection{Generation of light and charge}

For a given energy deposit $E_u$ of an interaction of type $u$ ($u \equiv \n{nr}$ for nuclear recoil or $u \equiv \n{ee}$ for electronic recoil)  in the presence of a drift field of strength \Efield, the  combined probability $ P(\Nph,N_e\,|\,E_u,\Efield)$ for the generation of \Nph\ photons and $N_e$ {escaping} electrons shows, in general, an anti-correlation between the number of {photons and electrons}. This is due to charge recombination processes which can lead to additional scintillation light\,\cite{Schwentner85}. At low
energy deposits, such as for nuclear recoils in the dark matter region of interest, however, the measurement
uncertainties {due to  statistical fluctuations in the number of generated photons}  dominate the width of the observed probability distributions~\cite{EDahlPhD}. Hence, $P$ can be
approximated by independent Poisson processes:

\begin{align}
 P(\Nph,N_e\,|\,E_u,\Efield) &\approx   \Poi(\Nph \| \nph) \; \Poi(N_e \| n_e).
\label{e:indep:gen} 
\end{align}

The average energies needed for creation of one information carrier (photon or free electron) are expressed by effective ``$W$-values'', which depend on the interaction type, the drift field, and, at low energies, also on the deposited energy. The field dependence can be factorized with functions $S_u$ for the reduction in light
yield due to field quenching and $T_u$ for the loss of charge due to recombination.
The expectation values $\nph(E_u,\Efield)$ and $n_e(E_u,\Efield)$ can then be written as

\begin{align}
\nph(E_u,\Efield)&=\frac{E_u}{\Wph(E_u,\Efield)} \approx \frac{E_u}{\Wph(E_u,\Efield=0)} \: S_u(\Efield),   \label{eq:n_gamma} \\
n_e(E_u,\Efield)&=\frac{E_u}{W_e(E_u,\Efield)} 
\approx\frac{E_u}{W_e(E_u,\Efield\rightarrow\!\infty)} \,T_u(\Efield),  \label{eq:n_e} 
\end{align}
where $S_u(\Efield=0) = 1$ and $T_u(\Efield\rightarrow\infty) = 1$.

In xenon dark matter detectors, the energy calibration of nuclear recoils is accomplished by comparing the signals from known $\gamma$-ray lines
to dedicated measurements of the functions $\Wph$ and $S_u$, or $W_e$ and $T_u$, which differ for electronic and nuclear recoils.  In order to minimize the systematic uncertainties of the cross-calibration resulting from modelling the detector responses, a reference source is frequently used. Historically, this has been the 122\,keV$_{\n{ee}}$ line from $^{57}$Co decay, where  the keV$_{\n{ee}}$ represents the electronic-equivalent recoil energy.  The in situ measured  light and charge yields at 122\,keV$_{\n{ee}}$ can be used as fixed points to establish the energy scale at lower $\gamma$-ray energies and for nuclear recoils, using the ratios of $\Wph(E_u)$ and $W_e(E_u)$ relative to this reference, and applying the functions $S_u(\Efield)$ and $T_u(\Efield)$, respectively. Currently, for nuclear recoils ($E_u \equiv E_{\n{nr}}$),  $\Wph(E_{\n{nr}})$~\cite{Plante:2011hw} has been measured to lower energies than $W_e(E_{\n{nr}})$.  Hence, XENON100 uses the primary scintillation light to establish the energy scale {for nuclear recoils}.

\subsubsection{Measurement of the primary scintillation light} %- - - - - - - -

The expectation value for the primary scintillation light signal S1 on PMT $i$  with a gain $g_i$ in units of electron charge is
\begin{align}
\textrm{s}1_{i}^{q}(\vec{r})  &= \nph(E_u,\Efield) \gamma_i(\vec{r})· \eta_i\: g_i =\nph(E_u,\Efield) \mu_i(\vec{r})\: g_i,
\end{align}
where $q$ denotes an integral over the current pulse, $\gamma_i(\vec{r})$ is the probability for a photon created at position $\vec{r}$ within the TPC
 to reach the photocathode of PMT $i$ and $\eta_i$ is the product of quantum and collection efficiencies for that  PMT.  The combined function $\mu_i(\vec{r})$ is the light detection efficiency and is measured with $\gamma$-ray calibrations (see Sec.~\ref{EventRec}).  The raw data processor converts the measured signal into units of photoelectrons, using the estimates of gains which include an electronic amplification factor of $10$~\cite{Aprile:2011dd}. 
Since errors in the determination of the combined PMT and amplification gains are typically $ < 2\%$~\cite{Aprile:2011dd}, these will be neglected in the following. Hence the expectation value for the primary scintillation signal on PMT $i$ in units of photoelectrons (PE) is

\begin{align} \label{e:sone:i}
\textrm{s}1_i(\vec{r})   & 
\approx \nph(E_u,\Efield) \: \mu_i(\vec{r}).
\end{align}

\noindent
If $M$ is the number of PMTs in the TPC, the energy deposit  for nuclear recoils $E_{\n{nr}}$ determines the expected total primary scintillation signal s1 as

\begin{align}
\textrm{s}1(\vec{r}) =& \sum_{i=1}^M \textrm{s}1_i(\vec{r})  \approx \nph(E_{\n{nr}},\Efield) \: \mu(\vec{r}) 
\nonumber \\
= & E_{\n{nr}} \: {\cal L}_y(E_{\n{ee}}=E_\mathrm{ref},\Efield,\vec{r}) \: \nonumber\\ 
&\times \Leff(E_{\n{nr}}, \Efield=0) \: \frac{S_{\n{nr}}(\Efield)}{S_{\n{ee}}(\Efield)},
 \label{e:s1:Leff}
\end{align}
\noindent
where $\mu(\vec{r}) = \sum_i \mu_i(\vec{r})$. ${\cal L}_y$ is the measured light yield (in PE/keV$_{\n{ee}}$) for a reference  $\gamma$-ray energy $E_\mathrm{ref}$ at the given electric field  and position [see Eqs.~(\ref{eq:n_gamma}) and (\ref{e:sone:i})], given as 

\begin{align}
{\cal L}_y(E_{\n{ee}}=E_\mathrm{ref},\Efield,\vec{r}) = \frac{S_{\n{ee}}(\Efield)\,\mu(\vec{r})}{\Wph(E_{\n{ee}}=E_\mathrm{ref},\Efield=0)}.
\end{align}
$\Leff$ is the relative scintillation yield of nuclear recoils with respect to the reference $\gamma$-ray line at zero field, which equals
\begin{align}
\Leff(E_{\n{nr}},\Efield=0) = \frac{\Wph(E_{\n{ee}}=E_\mathrm{ref},\Efield=0)}{\Wph(E_{\n{nr}},\Efield=0)}.
\end{align}
$S_{\n{nr}}$ and $S_{\n{ee}}$ are the reductions in light yield due to field quenching 
for nuclear and electronic recoils, respectively. 

The data analysis is usually performed with the ``spatially-corrected" measured signal 

\begin{align}
\textrm{cS1} &\equiv  \textrm{S1}(\vec{r})  \frac{ \langle \mu \rangle } {\mu (\vec{r})}, \label{cS1}
\end{align}
\noindent
where $\langle \mu \rangle$ is the spatial average. For this spatially-corrected signal, cs1 corresponds to the detector-averaged signal expectation value

\begin{align}
\textrm{cs1} &\approx \nph(E_{\n{nr}},\Efield)  \langle\mu \rangle \nonumber \\ &=  E_{\n{nr}} \: \langle{\cal L}_y (E_\mathrm{ref},\Efield)\rangle\: \Leff(E_{\n{nr}}, \Efield=0) \:\frac{S_{\n{nr}}(\Efield)}{S_{\n{ee}}(\Efield)},
\label{e:cs1:Leff}
\end{align}
\noindent
where $\langle{\cal L}_y\rangle$ is the detector-averaged light yield. 

Assuming a Poisson-distributed number of generated photons $\Nph$ and a binomially distributed number of generated photoelectrons for each  PMT $\Nei$, the pdf is described by 

\begin{align}
p_{\textrm{S1},i}(\textrm{S1}_i\!\|\!\nph(E_u,\Efield)) \, d\textrm{S1}_i =&\!\sum_{\Nei}\!\sum_{\Nph} \!\probpmti (\textrm{S1}_i \|\Nei)\,d\textrm{S1}_i \nonumber \\
  &\times \Binom  (\Nei \| \Nph,\mu_i(\vec{r}))& \nonumber \\
 &\times \Poi(\Nph \| \nph(E_u,\Efield)) \nonumber \\
                    =&\sum_{\Nei} \probpmti (\textrm{S1}_i \| \Nei) \nonumber \\
			&\times \Poi(\Nei \| \nph\,\mu_i(\vec{r}))   \, d\textrm{S1}_i,
\label{e:sonepmt:pdf}
\end{align} 
where $\probpmti (\ldots)$ is the response of PMT $i$, approximated  by a Gaussian with mean value $\Nei$ and width $\sigma_{\textrm{PMT}_i} \sqrt{\Nei}$. The detected photoelectron width $\sigma_{\textrm{PMT}_i}$  is determined from PMT calibrations.

Taking into account that $\textrm{S1} = \sum_{i=1}^M \textrm{S1}_i$ the pdf for the total light signal S1 can be calculated from Eq.~(\ref{e:sonepmt:pdf}) as

\begin{align}
p_{\textrm{S1}}(\textrm{S1}\!\|\!\nph(E_u,\Efield)) \, d\textrm{S1}  =&  \nonumber \\
=&\left(\,\idotsint\limits_{1 \ldots\, M} \right. \prod_{i=1}^{M} 
        p_{\textrm{S1},i}(\textrm{S1}_i \| \nph) \times  & \nonumber \\
  &\delta(\textrm{S1}-\!\left.\!\sum_{j=1}^{M} \textrm{S1}_j) \:d\textrm{S1}_1\!\ldots\!
     d\textrm{S1}_{M}\!\right)\! d\textrm{S1}.&
\label{e:sone:pdf:long}
\end{align}
This formula can be evaluated by computer simulations using individual response tables for each PMT. 
When considering only one spatially-averaged detector response where all PMTs are described by the same average response, Eq.~(\ref{e:sone:pdf:long}) simplifies to

\begin{align}
p_{\textrm{cS1}}(\textrm{cS1} \|\nph(E_u,\Efield)) \, d\textrm{cS1}  \approx &\sum_{N_{pe}} \probpmt(\textrm{cS1} \| N_{pe}) \nonumber \\
&\times \Poi(N_{pe} \| \langle\mu \rangle \nph) \, d\textrm{cS1},
\label{e:sone:cs1:pdf:short}
\end{align}
where $N_{pe} = \sum_i \Nei$ is the sum over all photoelectrons released by the PMT photocathodes.
$N_{pe}$ is  Poisson-distributed with the spatially-averaged mean value cs1=$\nph\langle\mu \rangle $.
 $\probpmt(\textrm{cS1}|N_{pe})$ is the average PMT  response function, approximated by a Gaussian with mean value $N_{pe}$ and width $\sigma_{\textrm{PMT}}\sqrt{N_{pe}}$. The average single-photoelectron resolution of XENON100 PMTs $\sigma_{\textrm{PMT}}$ is {determined} to be  $\sigma_{\textrm{PMT}}=0.5$\,PE  {using PMT gain calibration data (see Fig.\,14 in~\cite{Aprile:2011dd})}.

\subsubsection{Measurement of the charge}% ----- - - - - - - ----

The ionization electrons produced at an interaction point drift through the liquid, where losses occur due to
attachment to electronegative impurities with characteristic time $\tau_e$  (see Sec.~\ref{EventRec}). 
Electrons reaching the liquid surface are extracted into the gas phase with a yield $\kappa$ that depends on the extraction field $\Efield_\mathrm{gas}$. The same field is responsible for the proportional scintillation light signal S2\,\cite{Dolgoshein} for which the light amplification factor $Y$ results from collisional excitation of the atoms in the gas by the field-accelerated electrons. The expectation value for this secondary scintillation light signal on PMT $i$ in
units of PE is described by

\begin{align}
s2_i(\vec{r})   &\approx  n_e(E_u,\Efield)\,e^{-t_d/\tau_e} \kappa(\Efield_\mathrm{gas}) \,Y\left( \frac{\Efield_\mathrm{gas}}{\rho},h_\mathrm{g}\right) \,\beta_i(x,y) \eta_i.
\label{eq:s2i:long}
\end{align} 
$Y$ is also called secondary scintillation gain and it depends on the ratio of $\Efield_\mathrm{gas}$ to the gas density $\rho$, and on the size of the gas gap $h_\mathrm{g}$. Due to mesh warping or to an inclined liquid level, $\Efield_\mathrm{gas}$ and $h_\mathrm{g}$ can be $(x,y)$ position dependent.
$\beta_i(x,y)$ is the probability  for a photon created at the  position $(x,y)$ in the gas gap to reach the photocathode of the PMT $i$.  Since the S2 signal is created in a narrow gas gap ($h_\mathrm{g} \sim 2.5$\,mm)
, $\beta_i(x,y)$ can be considered as a function of $(x,y)$ only.  Gamma calibration lines  can be used to measure the product $\delta_i = \kappa Y\beta_i\eta_i$. Currently, only the sum over the PMTs  is measured, resulting in an estimate $\delta(x,y) = \sum_i \delta_i(x,y)$. 

The analysis is usually performed with the corrected measured signal:
 
\begin{align}
\textrm{cS2}\equiv  \textrm{S2}(\vec{r})  \: e^{t_d/\tau_e} \: \frac{\langle \delta\rangle}{\delta(x,y)}.
\label{cS2}
\end{align}
Using Eqs.~(\ref{eq:n_e}) and (\ref{eq:s2i:long}), the expected total secondary scintillation signal for nuclear recoils can be written as

\begin{align}
s2(\vec{r})  = \sum_i s2_i(\vec{r}) = E_{\n{nr}}\:Q_y(E_{\n{nr}})\: e^{-t_d/\tau_e} \: \delta(x,y),
\label{e:stwo:tot}
\end{align} 
where $Q_y = T_{\n{nr}}(\Efield)/W_e(E_{\n{nr}},\Efield_\mathrm{ref})$ is the measured charge yield of nuclear recoils (in e$^-$/keV$_{\n{nr}}$) at the given electric field. 
The pdf $p_{\textrm{S2},i}(\textrm{S2}_i)$ can be described as

\begin{align}
p_{\textrm{S2},i}(\textrm{S2}_i| n_e(E_u,\Efield)) \, d\textrm{S2}_i =& \sum_{\Nei}  \probpmti (\textrm{S}2_i \| \Nei) \nonumber \\
			&\times \Poi(\Nei \| n_e, \delta_i)   \, d\textrm{S}2_i,
\end{align} 
where $\probpmti$ is the same response of PMT $i$ as in Eq.~(\ref{e:sonepmt:pdf}). The pdf for the total proportional scintillation light signal S2, denoted by $p_{\textrm{S2}}(\textrm{S2}| n_e(E_u,\Efield))$, can be evaluated analogously to Eq.~(\ref{e:sone:pdf:long}).

\subsubsection{Event rate calculation} % - -- - - - - - - - -

The measured differential nuclear recoil rate 
$d^2R/d\textrm{S1}\,d\textrm{S2}$  for a given WIMP-nucleus scattering rate  $dR/d\textrm{E}_{\textrm{nr}}$ is computed as

\begin{align}
\frac{d^2R}{d\textrm{S1}\,d\textrm{S2}} = &\,\epsilon(\textrm{S1,S2}) \:  \int  \dRdE  \:  p(\textrm{S1},\textrm{S2}\,|\,E_{\n{nr}}) \, dE_{\n{nr}}  \nonumber \\
		\approx &\,\epsilon_{\textrm{1}}(\textrm{S1})\,\epsilon_{\textrm{2}}(\textrm{S2}) \nonumber \\
		 &\times  \!\int\!\dRdE p_{\textrm{S1}}(\textrm{S1}|E_{\n{nr}})\, p_{\textrm{S2}}(\textrm{S2}|E_{\n{nr}})  \, dE_{\n{nr}},
\end{align}
where $\epsilon(\textrm{S1,S2})$ is the two-dimensional and $\epsilon_{\textrm{1}}$(S1) and $\epsilon_{\textrm{2}}$(S2) are one-dimensional signal detection efficiencies for the applied selection criteria on the data. These are a combination of the trigger threshold, event search algorithm and event selection cuts, where cuts can be applied on both spatially corrected  and uncorrected S1 and S2 signals.

As already mentioned, so far the relation between  $E_{\n{nr}}$  and S1 signal has been measured more precisely and down to lower energies ($\nph(E_{\n{nr}},\Efield)$ via $\Leff(E_{\n{nr}})$)  than the relation between $E_{\n{nr}}$ and S2 signal ($n_e(E_{\n{nr}},\Efield)$ via $Q_y(E_{\n{nr}})$; see Fig.\,5 of reference~\cite{Bezrukov:2010qa}). Therefore, the analysis is done using only the measured differential rate expressed in the cS1 signal as

\begin{align}
\frac{dR}{d\textrm{cS1}} 
  \approx &\,\epsilon_{\textrm{c1}}(\textrm{cS}1) \int  \dRdE  \: p_{\textrm{cS1}}(\textrm{cS1} \| E_{\n{nr}})
 \nonumber \\		 &\times\underbrace{\int\limits_{\textrm{S2}_{\textrm{min}}}\!\epsilon_{\textrm{2}}(\textrm{S2})\,p_{\textrm{S2}}(\textrm{S2}|E_{\n{nr}}) \, d\textrm{S2}}_{\epsilon_{\textrm{E2}}(\textrm{E}_{\n{nr}} \propto \textrm{cs1})} \: dE_{\n{nr}} \nonumber \\
   = &\,\epsilon_{\textrm{c1}}(\textrm{cS}1) \int  \dRdE \, \epsilon_{\textrm{E2}}(\textrm{E}_{\n{nr}}) \, p_{\textrm{cS1}}(\textrm{cS1} \| E_{\n{nr}})\, dE_{\n{nr}},
  \label{eq:drdcs1}
\end{align}
where for most of the selection criteria the efficiency is estimated in the spatially-corrected S1 signal $\epsilon_{\textrm{c1}}(\textrm{cS1})$.  The exception is the S2 threshold cut ($\textrm{S2}_{\textrm{min}}$), which  directly influences the corresponding S1 signal efficiency via energy sharing at the level of generated $\nph$ and $n_e$ before the Poisson fluctuations and the measurement uncertainties. Hence its acceptance $\epsilon_{\textrm{E2}}$ is calculated as a function of  the detector-averaged expectation value cs1 (see Sec.~\ref{CutE}), which is related to nuclear recoil energy $E_{\n{nr}}$ via Eq.~(\ref{e:cs1:Leff}). 
For notation simplicity, in the rest of the paper  we use S1 and S2 instead of the corrected quantities cS1 and cS2, unless explicitly specified.

%%%%%%%%%%%%%%%%%%%%%%%%%%%%%%%%%%%%%%%%%%%%%%%%%%
\section{Data analysis}
\label{chap::ScienceRun}
%%%%%%%%%%%%%%%%%%%%%%%%%%%%%%%%%%%%%%%%%%%%%%%%%%

This section presents the analysis associated with {two dark matter search runs acquired in the period 2010--2012.} {The first science run, called Run-I here, was acquired in 2010 and provides 100.9\,live days of data. The second science run, Run-II, was recorded in 2011 and 2012 after an extensive distillation campaign to remove krypton in order to reduce its radioactive isotope $^{85}$Kr (see Sec.~\ref{Background}) and has 224.9\, live days of data. Both data sets have been used to obtain results which have}  already been interpreted in terms of spin-independent elastic\,{\cite{Aprile:2011hi,Aprile:2012kx}, inelastic\,\cite{Aprile:2011ts} and spin-dependent elastic\,\cite{Aprile:2013doa}} WIMP-nucleon interactions. {The two data sets, analysis methods, event selection and acceptances, and the background predictions are described in detail below. The main characteristics of Run-I and Run-II are summarized in Table~\ref{tab:Dataset}. The two runs share many other attributes; where there are minor differences, they are reported in the text and in parenthesis for Run-I and Run-II, respectively.}

\begin{table}[h]
\caption{Key parameters of the two science runs.}
\label{tab:Dataset}
\begin{center}
\begin{tabular}{lcc}
\hline
 & Run-I & Run-II \\
\hline
\hline
Livetime [days] & 100.9 & 224.6\\
%Run start & 13/01/2010 & 28/02/2011\\
%Run end & 08/06/2010 & 31/03/2012\\
Run start & \ \  Jan 13, 2010 & \ \ Feb 28, 2011\\
Run end & Jun 8, 2010 & Mar 31, 2012\\
%Outer Rn level [Bq/m$^3$] &350& 300\\
Anode voltage [kV] & +4.5 & +4.4\\
Cathode voltage [kV] & $-$16 & $-$16\\
DAQ deadtime & 10\% & $<$1\% \\
\hline
\end{tabular}
\end{center}
\end{table}

\subsection{Data sets} % =======================================

\begin{figure}[h]
\centering
\includegraphics[width=1\columnwidth]{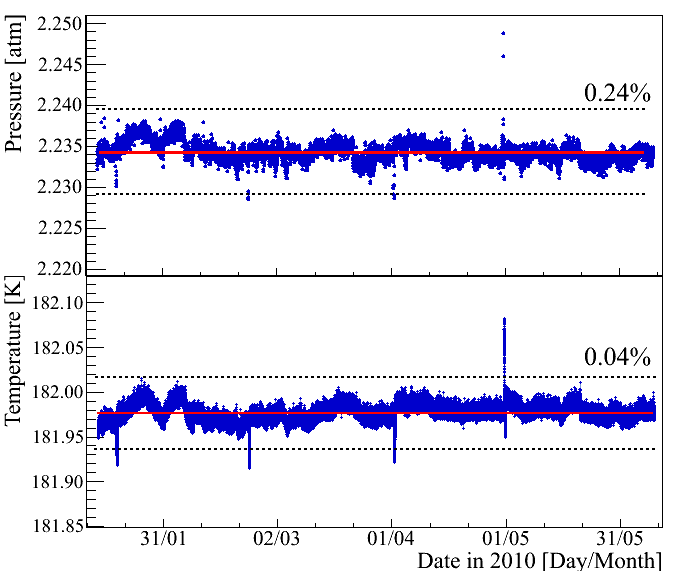}
\caption{Pressure and temperature of the XENON100 detector during {Run-I}. The dotted lines represent the maximum allowed variations{, which are quantified by the numbers,} and periods falling outside are removed from the dataset. {A similar analysis is provided in Ref.~\cite{Aprile:2012gk} for Run-II.}}
\label{fig:PT}
\end{figure}

The data used in {the analyses} were selected from periods with stable detector operating conditions. Periods with xenon pressure and temperature values being more than five sigma away from the average value were excluded from the analysis. After this selection, other parameters such as LXe level, cryostat vacuum pressure, purification flow and pulse tube refrigerator temperature were found to be stable during the whole science run. {For example, Fig.}~\ref{fig:PT} shows the time evolution of the xenon pressure and temperature over the entire length of {Run-I}, including correlated fluctuations. A 5-sigma  variation corresponds to $\pm \,0.04 $\,K  ($0.04\%$) and $\pm  \,0.005\1{atm}$ ($0.24\%$) in temperature and pressure, respectively. 

The high voltages biasing the {cathode and anode} were continuously monitored. The cathode was stable over {the complete period of a given run while the anode had occasional trips}  due to the anode current exceeding a predefined threshold. Data from 20~minutes before the trip until 20~minutes after restoration of {the anode voltage} were removed from the analysis, and the livetime was corrected accordingly. About 2\%{(1\%)} of the data were removed due to variations in the detector's operation parameters.

The radon concentration in the XENON room and inside the shield cavity were measured using dedicated radon monitors\footnote{RAD7 from Durridge.}. The radon concentration in the XENON room was about
{$320$\,Bq/m${^3}$.} The volume inside the shield cavity was constantly flushed with boil-off nitrogen gas {to reduce the radon induced background.}  The measured electronic recoil background rate was found to be stable over the whole time in various energy intervals. {Periods of 18\,(7)~\,live~days were} removed from the data due to an increased level of electronic pick-up noise. 

The PMT gains were monitored with weekly LED calibrations and found to be stable within 2\% which is given by the precision of the measurement. Out of the 242 PMTs installed,  4 in the top array, 4 in the bottom array and 5 in the veto were non-functional throughout {the runs}. Weekly $^{137}$Cs calibration data ($\gamma$-ray line at 662\,keV$_{\n{ee}}$) were used to study the evolution of the light yield, charge yield, LXe purity, as well as the width of S2 pulses. The data was stable during { both runs} and the LXe purity was continuously increasing. {Taking into account time periods removed due to the above-mentioned instabilities, anode trips, high electronic noise levels and DAQ dead time, the total live time used for the analysis of Run-I was 100.9\,live days, while it was 224.6\, live days for Run-II.}

\subsection{Data acquisition and calibration} % - -- - - - - - - - -

\begin{figure}[!h]
\centering
\includegraphics[width=1\columnwidth]{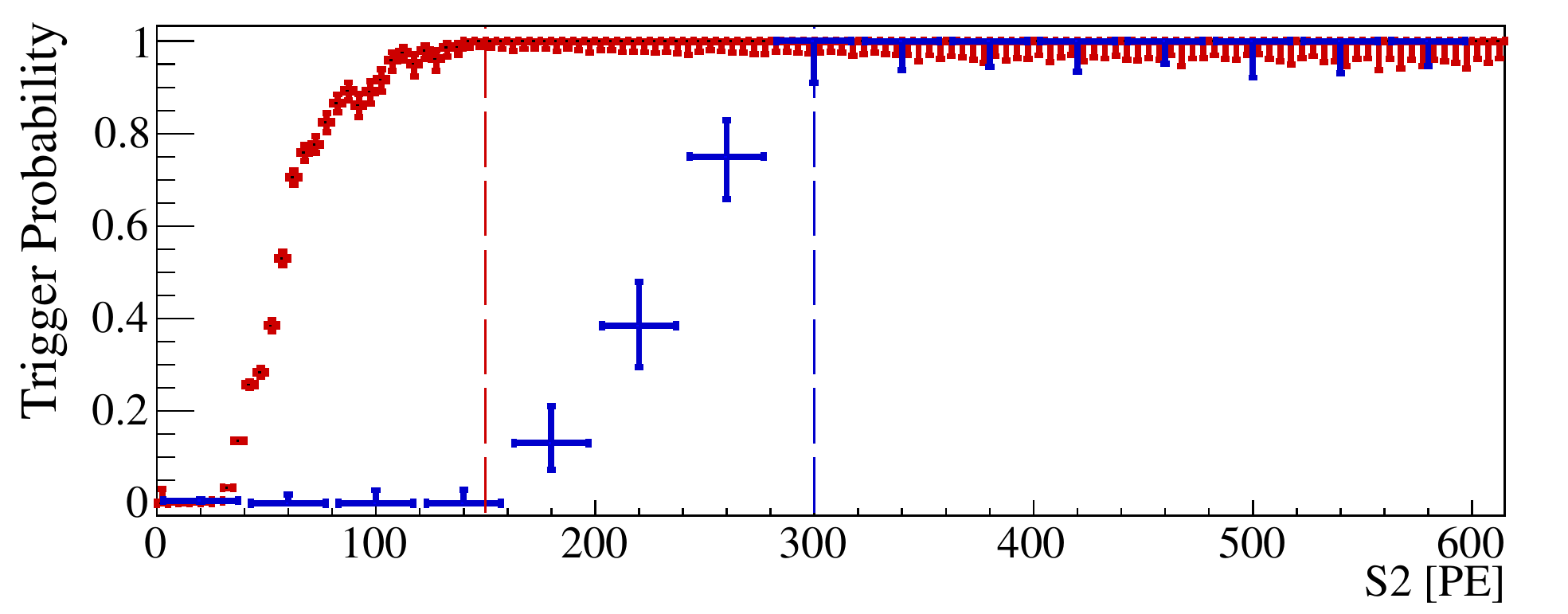}
\caption{Directly measured trigger probability as a function of the S2 signal size. It is unity above { the 300\,PE Run-I trigger threshold (blue) and 150\,PE Run-II trigger threshold (red).}}
\label{fig:trigger}
\end{figure}

{At low energies, the detector was triggered using the S2 pulses of typical 1\,$\mu$s spread. In Run-I, the analog sum of 68\,PMTs in the inner part of the top array and 16\,PMTs in the center of the bottom array was amplified, integrated, and shaped with a spectroscopy amplifier, passed through a low threshold discriminator, and distributed simultaneously to the ADCs~\cite{Aprile:2011dd}. For Run-II the trigger threshold was lowered by changing to a majority trigger mode: every PMT in the TPC exceeding a threshold of $\sim$0.5\,PE issued a voltage pulse of 125\,mV height. The sum of these voltage pulses was again shaped by the spectroscopy amplifier and fed into the discriminator to generate the logic trigger signal.

The trigger efficiency was 100\% for S2$>$300\,PE and S2$>$150\,PE for Run-I and Run-II, respectively. This was measured using three different methods~\cite{Aprile:2011dd} yielding consistent results: indirectly by observing the spectral roll-off from different sources, directly by using square-wave test pulses of a known size, and directly using real S2 pulses in dedicated background and calibration measurements. In the last method, the logical trigger signal was digitized using an empty ADC channel such that the information whether a peak generated a trigger or not was present for every S2 peak in the waveform. Small S2 peaks following the main one (triggering the digitization of the event),  see Fig.~\ref{fig:GoodEvents}, are mainly caused by photoionization of impurities in the liquid phase or TPC materials affected by photons from the S1 or S2 signal\,\cite{Santos:2011ui}. By comparing the spectrum of the secondary S2 peaks which raised a trigger to all S2 signals, one can derive the trigger functions as shown in Fig.~\ref{fig:trigger}. Since the measurement for Run-I was limited in statistics much more data was taken for Run-II.}

{
\begin{table}[h]
\caption{Summary of electronic recoil (ER) and nuclear recoil (NR, from an AmBe neutron source) calibration parameters including the electron lifetime (EL) for the neutron calibration and for the science data taking periods.}
\label{tab:Calibration}
\begin{center}
\begin{tabular}{lcc}
\hline
 & Run-I & Run-II \\
\hline
\hline
Livetime ER [days] & 5.8 & 48.0 \\
\hline
Livetime AmBe [days] & 2.9 & 2.7 \\
\hline
EL during AmBe [$\mu$s] & 210 & 360 \\
EL science data [$\mu$s] & (230 -- 380) & (374 -- 611) \\
\hline
\end{tabular}
\end{center}
\end{table}
}

The flat low-energy Compton continuum in $^{60}$Co calibration data were used to characterize the detector's response to electronic recoils {(ER)} during Run-I. The source was placed at three different positions around the detector to evenly cover the sensitive LXe volume of the TPC. {In Run-II, this data was augmented with calibration data from a $^{232}$Th source, extending all around the detector. The calibration data collection was spread in time over a given science run.}  

{The nuclear recoil (NR) region was defined using elastic neutron interactions from a $^{241}$AmBe source. The NR data were acquired in one dedicated measurement just before the start of Run-I and in two measurements right before and right after Run-II. Most of the data quality cuts and their acceptance have been defined on these data (see Sec.~\ref{Cuts}). Table~\ref{tab:Calibration} summarizes all the calibration data during the two science runs.} 

The irradiation of LXe with neutrons also gives $\gamma$-ray lines at 40\,keV$_{\n{ee}}$ and 80\,keV$_{\n{ee}}$ from $^{129}$Xe and $^{131}$Xe, respectively. The xenon isotopes are also activated leading to delayed de-excitation lines mainly at 164\,keV$_{\n{ee}}$ ($\tau_{1/2} = 11.8$\,d) and 236\,keV$_{\n{ee}}$ ($\tau_{1/2} = 8.9$\,d) from $^{131m}$Xe and $^{129m}$Xe, respectively. Some of these lines are used to define position corrections for the S1 and S2 signals. 

The energy scale for  nuclear recoils $E_{\n{nr}}$ {was} inferred from the cS1 signal via Eqs.~(\ref{cS1}) and (\ref{e:cs1:Leff}). The scintillation efficiency $\Leff(E_{\n{nr}})$ of nuclear recoils relative to 122\,keV$_{\n{ee}}$ {was} parametrized using all existing direct measurements as shown in~\cite{Aprile:2011hi}. The scintillation quenching factors due to the applied electric field  for electronic recoils $S_{\n{ee}}$ = 0.58 and nuclear recoils $S_{\n{nr}}$ = 0.95 were taken from~\cite{Aprile:2006kx}. {For Run-I, the} light yield at 122\,keV$_{\n{ee}}$ was $\langle{\cal L}_y\rangle=(2.20\pm0.09)$\,PE/keV$_{\n{ee}}$  and was interpolated using the light yields from the 40\,keV$_{\n{ee}}$, 80\,keV$_{\n{ee}}$, 164\,keV$_{\n{ee}}$ and 662\,keV$_{\n{ee}}$ lines~\cite{Aprile:2011dd}. {The light yield increased to $\langle{\cal L}_y\rangle=(2.28\pm0.04)$\,PE/keV$_{\n{ee}}$ in Run-II. The energy interpolation was performed in this run using the NEST~\cite{Szydagis:2011tk} model.}

The science data in the nuclear recoil region was blinded and therefore not accessible until the analysis was finalized.  {The} lower 90\% quantile of the S2/S1 discrimination parameter for electronic recoils determined the blinding cut for energies S1$<160$\,PE {(S1$<100$\,PE)}. The data outside of the blinded region was also used for the estimation of the electronic recoil background in the WIMP signal region (see Sec.~\ref{Background}), for monitoring of the rate stability as well as for the estimation of the acceptances of certain data quality cuts.  

\subsection{Data processing}\label{EventRec} %--- --- --- --- --- 

The digitized waveforms from each of the 242 PMTs used in XENON100 are recorded. The trigger is located in the middle of the {400\,$\mu$s long waveform} and its global time is saved for each event. The raw data processor is based on the ROOT analysis toolkit\,\cite{Antcheva:2009zz}. It utilizes the difference in the S1 and S2  peak width, where the former is of the order of  a few tens of ns (see Fig.~\ref{fig:lowwidthcut}) mainly governed by the LXe scintillation decay time constants\,\cite{Hitachi83} and the latter is of the order of a few \,$\mu \textrm{s}$ determined by the electron cloud diffusion during the drift in the liquid and the drift time across the gas gap (see Fig.~\ref{fig:widthcut}). For each peak candidate found by the processor, several peak properties are calculated and stored, for example pulse area (in PE), height (in V), width (in ADC samples with 1\,sample = 10\,ns), etc. More details can be found in~\cite{Aprile:2011dd,GPlantePhD}.

For each identified S2, the $(x, y)$ position is calculated using the hit pattern of the S2 photons in the top PMT array. Three independent  algorithms were developed to reconstruct the position of each event. While the algorithm based on a neural network (NN)~\cite{ref:NN} is used for the analysis, the results from the two other algorithms (based on $\chi^2$ minimization and using support vector machines, SVM~\cite{SVM}) are used for quality cross checks, see {details in} ~\cite{Aprile:2011dd}. {A position resolution (1$\sigma$) of 3\,mm and 0.3\,mm are obtained in, respectively, $(x, y)$ and in $z$}. Because of the finite S2 signal width, only pulses which are more than 3\,mm apart in $z$ can be separated.

An energy deposit will produce a different number of detected S1 photons depending on its position in the TPC. Primarily, this is due to the different paths that the light travels until a PMT is hit. {For Run-I, a} 2D 
axial-symmetric S1 light collection map $\mu(r,z)$, determined from monoenergetic lines, was applied to each measured S1 light pulse to obtain the position-corrected value [see Eq.~(\ref{cS1})]. The maps derived using the 40\,keV$_{\n{ee}}$, 164\,keV$_{\n{ee}}$ and 662\,keV$_{\n{ee}}$ lines agree within 3\%. {The additional ER calibration statistics in Run-II allowed the 2D $\mu(r,z)$ light collection map to be replaced by a 3D $\mu(r,\theta,z)$ map improving the response close to the PMTs}. The maximum S1 correction for the whole target volume is less than a factor of~2 (see Fig.\,21 of~\cite{Aprile:2011dd}).

Similarly, there are position-dependent effects in the detection of the S2 signal [see Eq.~(\ref{cS2})]. A $(x, y)$ light collection map $\delta(x,y)$ is calculated, also from monoenergetic lines, using only the sum of S2 signals in the bottom PMT array (S2$_b$) as it has a more homogeneous light collection efficiency. The maximum $(x, y)$ correction in the {48\,kg fiducial mass, selected for the analysis of Run-I} (see~\ref{DiscrCus}), was 15\%{; it was a few percent smaller for Run-II due to the smaller fiducial volume.} An additional correction in $z$ is necessary due to absorption of ionization electrons by electronegative impurities in LXe during their drift. The  electron lifetime $\tau_e$, which is the time at which the total number of electrons produced is reduced by $1/e$, see Eq.~(\ref{eq:s2i:long}), is used to quantify this effect ({see Fig.\,19 of}~\cite{Aprile:2011dd}). The purity of the LXe was constantly increasing as monitored using the $^{137}$Cs full absorption peak due to the continuous purification of the xenon gas by a hot getter. {The electron lifetimes during the two science runs were compatible with the corresponding $^{241}$AmBe calibrations and continuously increased during the runs due to the increasing purity of the LXe. Their values at the beginning and the end of the runs are listed in Table~\ref{tab:Calibration}. The time evolution is shown in~\cite{Aprile:2011dd} for Run-I and in~\cite{Lavina:2013zxa} for Run-II.}

\subsection{Analysis methods}\label{subsec:AnalysisMethods} % ================

{
\begin{table}[h]
\caption{Summary of analysis parameters for the main Profile Likelihood (PL) analysis and for a second analysis, using a pre-defined signal search (benchmark) region. The benchmark analysis was used to directly compare the number of observed events to the one expected from background.}
\label{tab:AnalysisPars}
\begin{center}
\begin{tabular}{lcc}
\hline
 & Run-I & Run-II \\
\hline
\hline
PL ROI [PE] & (4 -- 30) & (3 -- 30)\\
PL ROI [keV$_{nr}$] & (8.4 -- 44.6)& (6.6 -- 43.3)\\

Benchmark ROI [PE] & (4 -- 30) & (3 -- 20)\\
Benchmark ROI [keV$_{nr}$] & (8.4 -- 44.6)& (6.6 -- 30.5)\\
S2 threshold [PE] & 300 & 150 \\
Benchmark ER discrimination & 99.75\% & 99.75\% \\
Benchmark NR lower contour & $\sim$3\,$\sigma$ & $\sim$97\% \\
Fiducial mass [kg] & 48 & 34 \\
\hline
\end{tabular}
\end{center}
\end{table}
 }

{For both science runs, it was decided before unblinding to use the Profile Likelihood analysis method introduced in\,\cite{Aprile:2011hx}, which does not employ a fixed discrimination in S2/S1 parameter space, as the primary interpretation method. A cuts-based analysis using a pre-defined $\log_{10}(\n{S2_b/S1})$ benchmark region for the WIMP search was also performed to cross check the result and to directly compare the observed number of events to the one expected from background. The region of interest (ROI) defines the energy interval used in a given analysis as inferred from the number of PE in S1 (see~\ref{CutE}). In the case of Run-I, the Profile Likelihood and the benchmark analysis shared the same ROI, while they were different for Run-II. Both the Profile Likelihood and benchmark analysis used the same S2 threshold specific to each run (see Sec.~\ref{DiscrCus}). Table~\ref{tab:AnalysisPars} summarizes the main analysis parameters for the two runs. The fiducial mass determination is described in~\ref{DiscrCus}.}

{For the Profile Likelihood analysis, the calibration data sets were divided into bands along $\log_{10}(\n{S2_b/S1})$ such that for every 1\,PE-wide S1 bin the signal-like events were equally distributed between the bands (see Fig.~\ref{fig:bands}). Each signal or background event has an associated probability to fall into a certain band. The number of bands was optimized based on the binning resolution of the discrimination parameter $\log_{10}(\n{S2_b/S1})$ and the available amount of calibration data in each band. While a large number of bands yields a result which depends less on the location of the event with respect to the band boundaries, the statistical uncertainty on the signal fraction in each band increases.  Twelve bands were found to be optimum and we verified that the results are robust to the number of bands, see also\,\cite{Aprile:2011hx}. 
Based on the event distribution for signal and background on these bands, both the signal-plus-background and the background-only hypotheses were tested regardless of the observed data. In the case of the exclusion of the signal hypothesis, the resulting limit on the WIMP-nucleon cross section is given at 90\% confidence level (CL).}

{For the secondary analysis, the benchmark region was constrained by the 99.75\% electronic recoil rejection line defined using ER data, the $\sim$3\,$\sigma$ ($\sim$97\% for Run-II) lower contour of the nuclear recoil distribution, the S2 threshold and the S1 ROI.}

\subsection{Dark matter search event selection}\label{Cuts} % ================

This section describes the criteria applied to the science data to select candidate events in the energy ROI, which fall into the following general categories: data quality cuts,  energy selection and threshold cut on S2, selection of single scatter events, consistency cuts, selection of the fiducial volume, and the signal region selection for the cut-based analysis.

The acceptance loss for each condition is determined using the fraction of events removed by this selection alone. 
For the case of independent selection cuts, this method gives a better estimate of the acceptance than using the fraction of the total events which pass the cut under study. The reason is that events which fail multiple cuts are far less likely to be valid events. For the cases in which the selection cuts were found not to be independent, their combined acceptance is derived. Visual inspection of a subset of both accepted and rejected event waveforms was used as a cross-check to validate the acceptance determination. The acceptance was evaluated using the energy range and the fiducial mass selected for the analysis (see~\ref{CutE} and \ref{DiscrCus}).
Nuclear recoil data {are} used to determine the  acceptance for most of the cuts,  with the few exceptions where conditions of neutron calibration datasets were found not to be representative for the full science data taking conditions.  

\subsubsection{Data quality}  % - -- - - - - - - - -

Data quality cuts remove non-physical or noisy waveforms. A two-fold PMT time-coincidence  within a 20\,ns time window  is required for a valid S1, where each PMT hit must have a signal size larger than 0.35\,PE.
This efficiently rejects PMT dark current signals as they are unlikely to happen simultaneously in multiple PMTs. Additionally, the S1 coincidence requirement was tightened  for the pulses where the S1 peak contains noisy PMT channels.  If known noisy PMTs contribute to the S1 coincidence, then the coincidence requirement is increased by the number of noisy PMTs\,\cite{Aprile:2011hi} to avoid identification of an S1 candidate with only one good PMT. {Throughout Run-I there were 8~noisy PMT channels, which were reduced to~3 in Run-II.}

\begin{figure}[h]
\centering
\includegraphics[width=1.0\columnwidth]{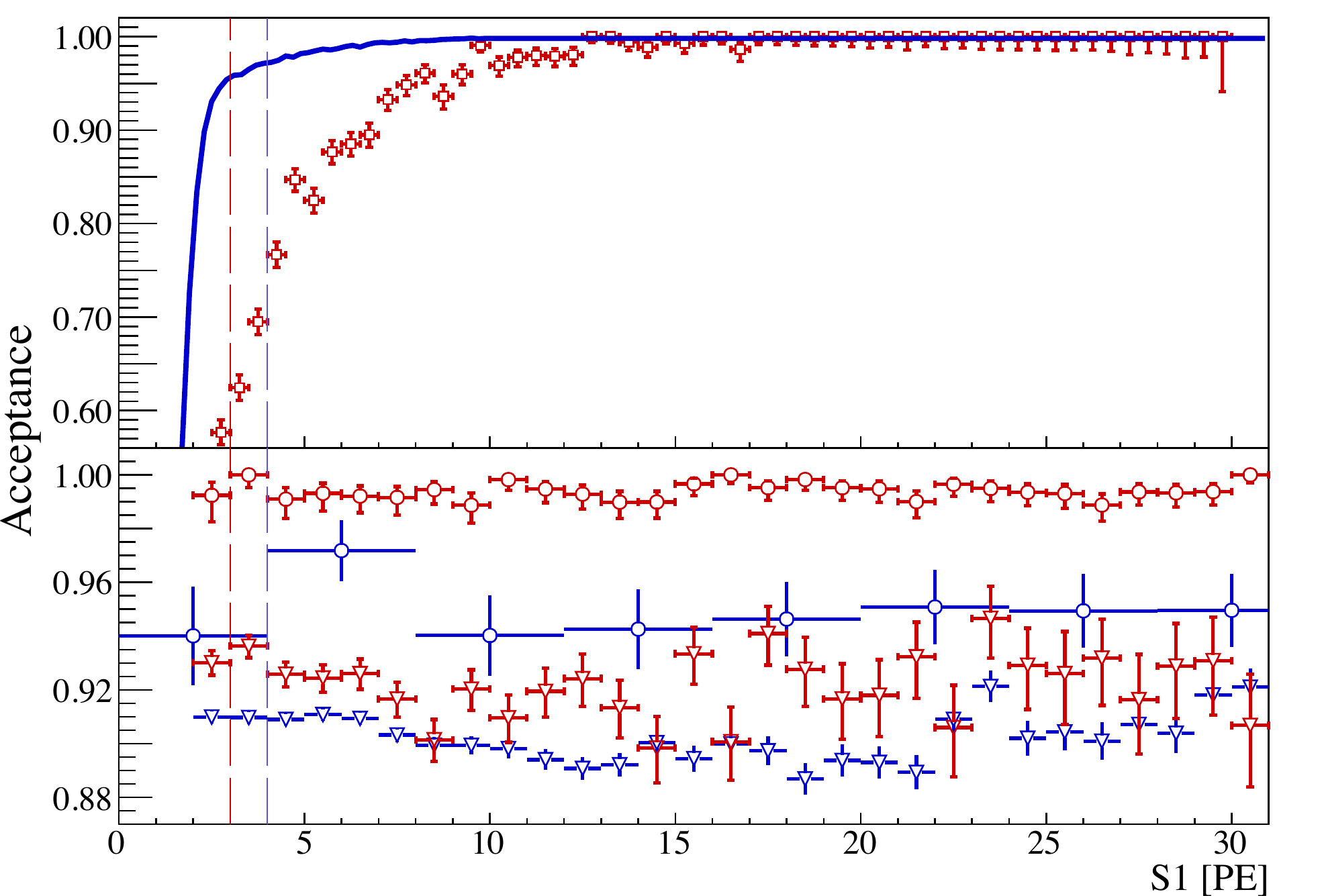}
\caption{{Acceptance of some of the cuts as a function of S1. Top: S1 coincidence requirement using the Run-I Monte Carlo (blue line) and the Run-II data-driven (open red squares) methods. Bottom:  S2 single scatter cut (circles) and S2 pulse width cut (triangles). Blue and red colors refer to Run-I and Run-II, respectively. The vertical dashed lines correspond to the lower energy threshold for Run-I (blue) and Run-II (red).} }
\label{fig:acc}
\end{figure}

For the {Run-I} results presented in\,\cite{Aprile:2011hi,Aprile:2011ts}, the acceptance {of the S1 coincidence cut} was computed via a Monte Carlo simulation, which takes into account light collection in the TPC, PMT quantum efficiencies, single photoelectron resolution, and the time dependence of the scintillation process. Using this method, the acceptance was estimated to be $>$\,99\% above 6\,PE and dropping down to 97\% at 4\,PE. However, the acceptance of this cut can be evaluated using data instead of a Monte Carlo simulation {and this improved method was employed in Run-II}. Events from nuclear recoil calibration data with a coincident signal in the veto {were} used to compute the acceptance of the coincidence requirement, as they correspond to true physical interactions in the TPC.  The acceptance { was} $>$\,99\% above 10\,PE and {dropped} down to 80\% at 4\,PE  (see Fig.~\ref{fig:acc}). {Using the lower acceptance at small S1 from the data-driven method for the analysis of \mbox{Run-I}} leads to exclusion limits which are about 8\% weaker above 100\,GeV/$c^2$, increasing to 30\% around 10\,GeV/$c^2$. This, however, has negligible impact on the interpretation of the results in\,\cite{Aprile:2011hi}. 
 
\begin{figure}[!h]
\centering
\includegraphics[width=1\columnwidth]{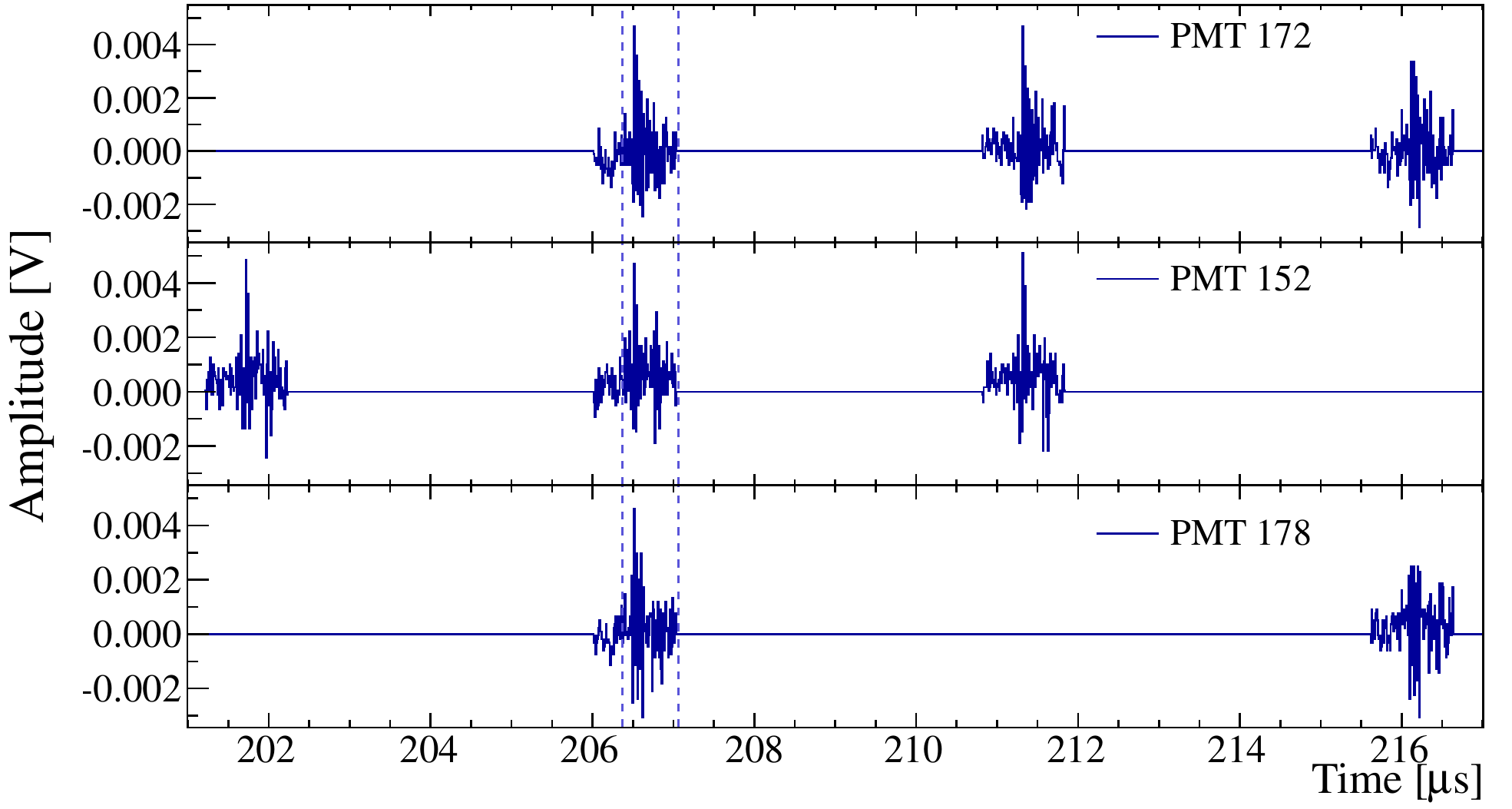}
\caption{PMT traces for a noise event found after unblinding {Run-I}. The S1 candidate region is indicated by blue dashed lines. All PMT signals contributing to the S1 candidate are periodic electronic pick-up noise peaks.}
\label{fig:NoiseEventPMT}
\end{figure}

{After unblinding Run-I, a} population of events with an S1 peak due to noise was found in the region around and below the analysis threshold of 4\,PE. This population { was} due to electronic pick-up noise and not related to physical events in the TPC. Since the original cuts were not sufficiently restrictive against electronic noise, two additional cuts had to be introduced post-unblinding in order to remove the full noise population, including three noise events with S1 signal leaking above 4\,PE.  
The S1 coincidence requirement was modified by {adding non-functional
PMT channels, that are turned off but can still pick-up electronic noise.} An example of a rejected noise event 
is shown in Fig.~\ref{fig:NoiseEventPMT}. The modification of the S1 coincidence requirement had no relevant impact on the nuclear recoil acceptance ($<0.4\%$ acceptance loss) as determined on $^{241}$AmBe calibration data. 

\begin{figure}[!h]
\centering
\includegraphics[width=1\columnwidth]{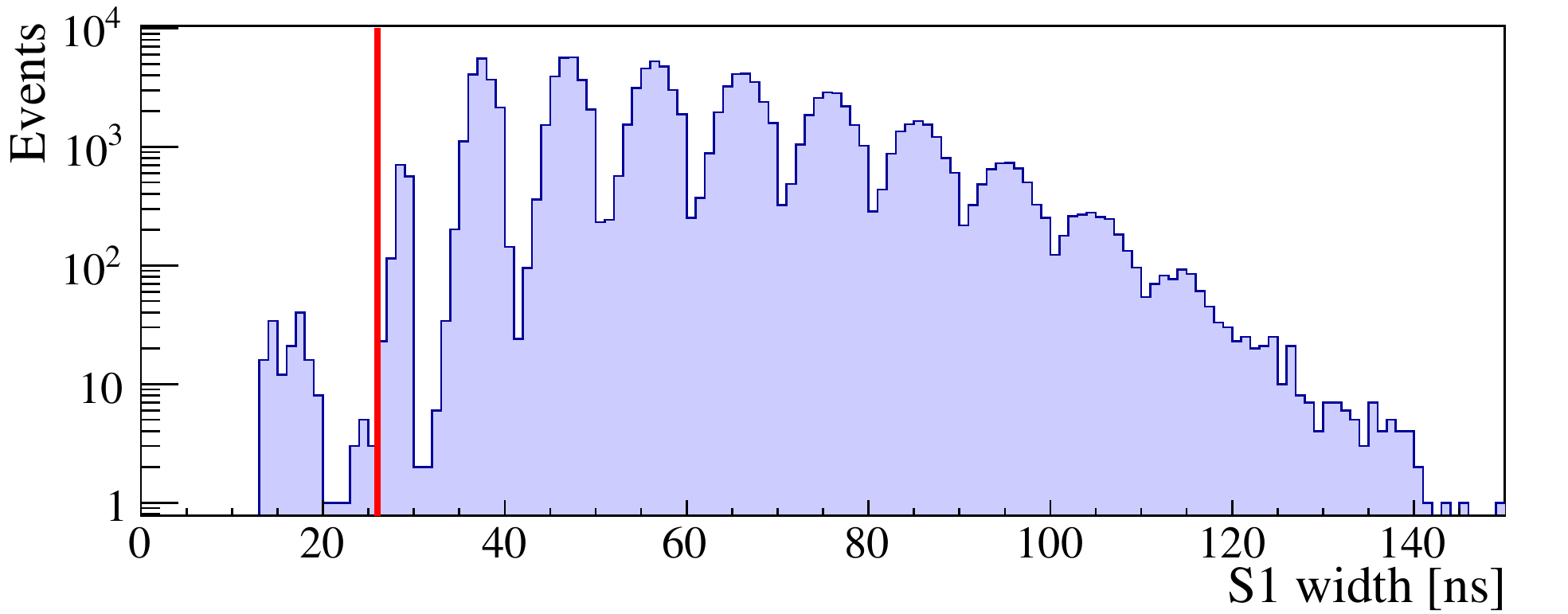}
\caption{Histogram of the S1 width at 10\% level for {NR data in Run-I}. The line represents the cut on this parameter, events left of the line are rejected to eliminate electronic noise.}
\label{fig:lowwidthcut}
\end{figure}

In addition, a cut on the S1~width was introduced. The peak width is defined as the distance from the left to the right peak boundary, determined by calculating the time where the trace exceeds 10\% of the peak height by linearly interpolating between the bins around this value.
Figure~\ref{fig:lowwidthcut} shows the histogram of the S1 peak width. 
The distribution has a regular structure due to the finite sampling frequency of 100\,MHz. Very small widths correspond to electronic noise pulses with samples fluctuating around the baseline, shortening the extent of the identified peak due to negative excursions (see Fig.~\ref{fig:NoiseEventPMT}).  Events left of the vertical line are rejected. The line was defined after visual inspection of waveforms from $^{241}$AmBe and low energy science data. Waveforms of valid events passing all cuts are shown in Fig.~\ref{fig:GoodEvents}. The S1 width cut has an acceptance of {97\% at 3\,PE,} 99.9\% at 4\,PE increasing to $>$99.9\% above 6\,PE, as verified with nuclear recoils.
{Both of these noise cuts were employed before unblinding in Run-II; no electronic pick-up noise events were observed in the Run-II event candidate list.}

There are events where no valid S1 peak is present in the waveform. Such events appear either if the real S1 is too small to be detected or is missed by the peak-finder, or if it is too close to the S2 peak to be resolved as an individual peak.  The latter mostly happens for interactions at the very top layer of LXe or in the gas region. In these events, a random electronic noise or a coincident PMT dark current peak can be picked up as an S1 candidate.  Their very asymmetric S2~signal distribution on the top and the bottom PMT arrays can be used to identify and reject them. This selection was defined using the non-blinded part of the science data.  {The cut acceptance was determined using neutron calibration data. It was unity above $\sim$10\,PE and decreased to $\sim$98\% at the analysis thresholds of the runs.}

The peak finding algorithm is tuned to correctly identify low-energy interactions with the highest possible acceptance. In waveforms of events with large energy depositions in the TPC (e.g. through-going muons) or with HV-induced micro-discharges, features in the pulse shape can sometimes be mistaken by the algorithm as individual small energy signals. The event selection requires that the integral of the largest S1 and S2 signals is bigger than the size of the integral of the remaining waveform.  In order to obtain a sample with DAQ conditions representative of the full science run, the acceptance of this cut was calculated using ER science data outside of the blinded region for Run-I and using the weekly high-statistics ER calibration data for Run-II. The cut  acceptance in the energy region of interest is about 98\%.

Occasionally,  S2-like events with signals very localized in the $(x,y)$ plane and/or unusual light patterns occur.  {In Run-I,} two classes of  such spurious events were found in some of the science data outside of the blinded region. In the first class, the events have the bulk of S2 light detected by a single PMT. As the maximal fraction of the S2 light seen by a single PMT has a mean value of about 20\,\%, events are rejected when this fraction is more than 65\%.
In the second class, events have an unusually small fraction of S2 light detected by the top PMT array. 
Since the S2 signal is generated in the gas phase, the top PMT array typically sees about 1.3 times more proportional scintillation light than the bottom array. Hence events are removed when the amount of S2 light detected by the top array is less than 74\% of the light in the bottom array.  Both cuts were optimized using  the non-blinded science data. Their acceptance was calculated using nuclear recoil data and it is above $99.6$\%. {These cuts were modified for Run-II in order to take into account the slightly different experimental conditions. The spurious events are rejected with high efficiency and a NR acceptance of unity by requiring that some of the S2 light must be seen by the top PMT array.}

\subsubsection{Energy selection and S2 signal threshold}\label{CutE}

{The S1 region of interest used in the analyses, together with the corresponding energies (see Eq.~\ref{e:cs1:Leff}), are listed in Table~\ref{tab:AnalysisPars}. The selection acceptances above 4\,PE were found to be high and nearly energy independent for Run-I, while for Run-II, improved noise conditions and the upgraded S2 trigger allowed reducing the S1 threshold to 3\,PE. No notable WIMP sensitivity was gained by increasing the S1 window above 30\,PE. The benchmark analysis in Run-II was limited to an upper threshold of 20\,PE chosen to optimize the signal-to-background ratio.}

\begin{figure}[!h]
\centering
\includegraphics[width=1\columnwidth]{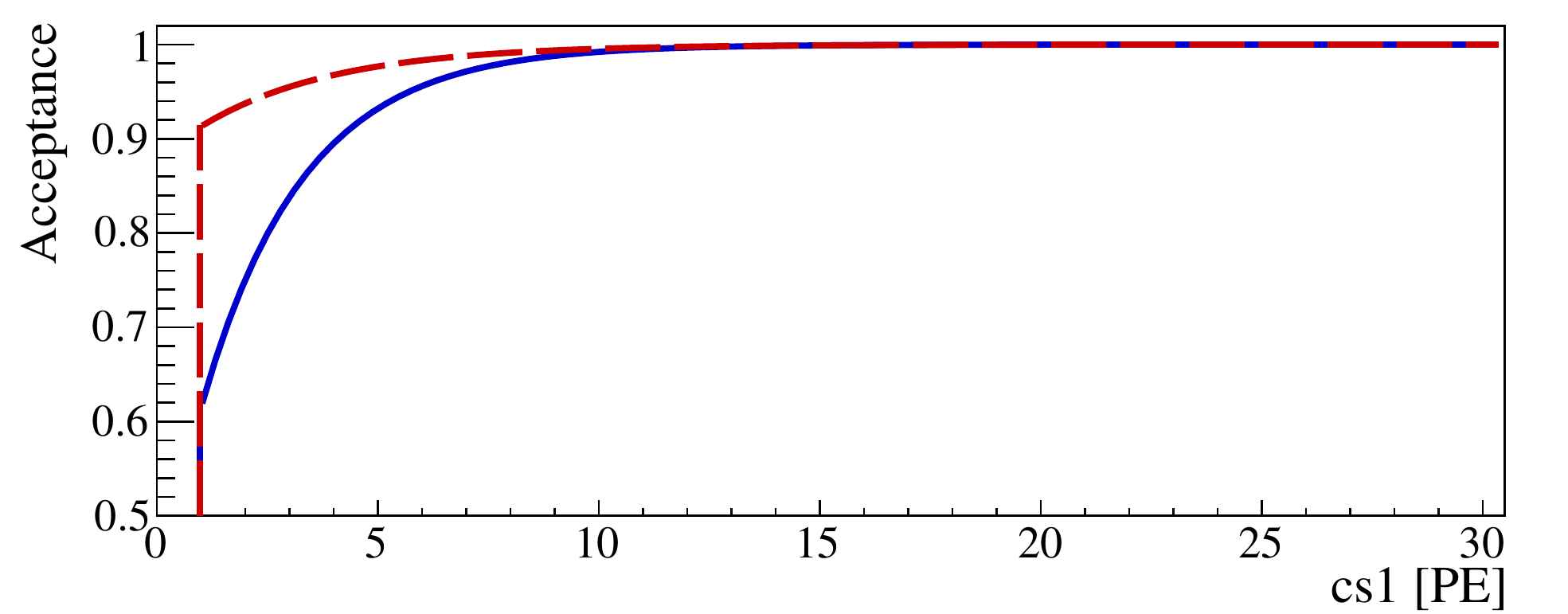}
\caption{Acceptance of the S2 threshold cut as a function of  cs1, 
the S1 signal before Poisson fluctuations {for Run-I (blue line) and Run-II (red dashed line)}. The sharp drop at 1\,PE was introduced artificially in order to be more conservative at the lowest signals. This acceptance is used in the WIMP rate calculation before the S1 signal is convoluted with a Poisson function to take statistical fluctuations into account.}
\label{fig:accS2thr}
\end{figure}

{For Run-I, in} order to stay well above the trigger threshold which starts to roll off around 280\,PE  (see Fig.~\ref{fig:trigger}) a valid event is required to have S2\,$>$\,300\,PE (uncorrected S2). {Because of the lowered trigger threshold, the condition was S2\,$>$\,150\,PE in Run-II.} The acceptance of such a condition affects the measured differential rate [see Eq.~(\ref{eq:drdcs1})] and is determined in the following way. For a given interval of  position-corrected cS1 from $^{241}$AmBe data, the position-corrected cS2 spectrum can be well described by a phenomenological function based on a continuous extension of the Poisson distribution

\begin{align}
f(\textrm{cS2}) = p_0 \times \Poi( \textrm{cS2}/p_1\| p_2 ), 
\end{align}
with three free parameters: $p_0$ to set the amplitude, $p_1$ to scale the cS2 axis, and $p_2$ to fit the mean value. $p_1$ may be kept constant, while $p_2$ varies approximately linearly with cS1.
The function is fitted to the data  and extrapolated conservatively to the range below threshold. In order to minimize the extrapolation, only the top 7.5\,cm of the detector {were} used for this analysis reducing the impact of charge losses due to the finite $\tau_e$. From this data the acceptance $\epsilon_{\textrm{c2}}(\textrm{cS1})$ of the {run-dependent S2 threshold} is determined numerically, taking into account the fiducial mass, the increasing $\tau_e$ during science data taking, and a spatially uniform event distribution as expected from WIMP interactions.

This acceptance, however, cannot simply be applied to the observed science data as the measurement-related fluctuations of the S1 and S2 signals are independent from each other (see Sect.~\ref{MeasurementProcess}). As an example, a recoil from a 7\,GeV/$c^2$ WIMP can enter the {analysis ROI} only because of upward fluctuations of the S1 scintillation signal. The average S2 signal of these events, however, is smaller than from WIMPs of higher masses, leading to a lower acceptance at the same S1 signal. The acceptance of the S2 threshold cut is therefore calculated as a function of the cs1 signal without Poisson fluctuations, which is proportional to the nuclear recoil energy deposition $E_{\n{nr}}$, see Eqs.~(\ref{e:s1:Leff})-(\ref{e:cs1:Leff}). Since the nuclear recoil spectrum varies with the WIMP mass, this leads to a WIMP-mass-dependent acceptance  (see Fig.\,2 of~\cite{Aprile:2011hi}). 

In general, cs1 and therefore the acceptance $\epsilon_{\textrm{E2}}(\textrm{cs1})$ are experimentally not accessible.  However, the latter can be determined indirectly from the previously derived $\epsilon_{\textrm{c2}}(\textrm{cS1})$ from $^{241}$AmBe data. 
Using $\Leff(E_{\n{nr}})$ from\,\cite{Aprile:2011hi} and Eq.~(\ref{e:cs1:Leff}), a $^{241}$AmBe Monte Carlo simulated energy spectrum is converted into a cs1 signal spectrum{, see~\cite{Aprile:2013teh} for details on the simulation}. An initially estimated acceptance function $\epsilon^*_{\textrm{E2}}(\textrm{cs1})$ is now applied to this spectrum and varied iteratively until the acceptance $\epsilon^*_{\textrm{c2}}(\textrm{cS1})$  from the Monte Carlo dataset convoluted with a Poissonian distribution to account for the statistical fluctuations is equal to 
 $\epsilon_{\textrm{c2}}(\textrm{cS1})$.
Figure~\ref{fig:accS2thr} shows the resulting {acceptances} $\epsilon_{\textrm{E2}}(\textrm{cs1})$ {for Run-I and Run-II. At low cs1, the acceptance is considerably higher in Run-II due to the lower S2 threshold of 150\,PE}. The acceptance is conservatively set to be 0 below 1\,PE.

\subsubsection{Selection of single scatter events}\label{singleScat}

{WIMPs are expected to scatter only once in the TPC producing one S1 and one S2 pulse in the waveform and no coincident signal in the veto volume. S2 pulses of a few PE size after the main S2 signal, e.g., small ionization signals from single electrons can bias the single scatter event selection as they can be mistaken as an S2 signal from a second scatter. It was observed that the size of these delayed signals is correlated with the size of the main S2 pulse. Thus, events with one S2 peak are selected by requiring all other S2 peaks to be smaller than a certain threshold value which was defined using nuclear recoil calibration data. The Run-I acceptance was calculated using ER science data, since its low-energy region was dominated by beta particle interactions, which represent an almost pure single scatter event sample. ER calibration data was used in the evaluation of the acceptance for Run-II.  The cut acceptance in the energy ROI is $\gtrsim$\,95\% for Run-I and $\gtrsim$\,99\% for Run-II (blue and red circles in Fig.~\ref{fig:acc}, respectively) and is independent of the S2 signal size. The better acceptance in Run-II is mainly due to the improved LXe purity leading to less background from photoionization and impurities.} 

Even in the case of an event with multiple scatters, only one S1 peak is expected since the 10\,ns sampling time does not allow to distinguish between them. Multiple S1 signals in a waveform would come either from pile up (which is very unlikely given the low trigger rate of \mbox{$\sim1$\,Hz} during science data taking), PMT dark current, electronic noise, or misidentified uncorrelated S2 ionization signals from single electrons. Events where a second S1 peak is seen by at least two PMTs are therefore removed if it appears within the  maximum possible drift time. However, if the S2/S1 ratio is too high for the second S1 candidate, the event is not rejected. {Since this cut is affected by the noise conditions during the measurement, the acceptance was determined from the ER science data in Run-I and from weekly high-statistics ER calibration data in Run-II. The cut acceptance in the ROI is $\sim$\,98.5\% for Run-I and $\sim$97\% for Run-II)}.

Given the negligible rate of accidental coincidences, a signal in the  active LXe veto which is in coincidence with the S1 peak in the TPC must therefore be from a double scatter interaction. All events where a coincident signal in the veto is present with a size $\geq 0.35$\,PE are rejected. The non-blinded science data {are} used to calculate the acceptance of the cut, which is { $\sim$99.5\% for both runs}.

\subsubsection{Consistency cuts} \label{ConsCuts}% --- --- --- ---

{Valid} events were selected based on the comparison of the reconstructed position estimators from different algorithms, the fit quality of the reconstructed position, and the comparison of the measured S1 PMT pattern with the one expected from the reconstructed position. Interactions in the gaseous xenon or events with  S1 and  S2 signals not from the same physical interaction were also removed.

 \begin{figure}[h]
\centering
\includegraphics[width=1\columnwidth]{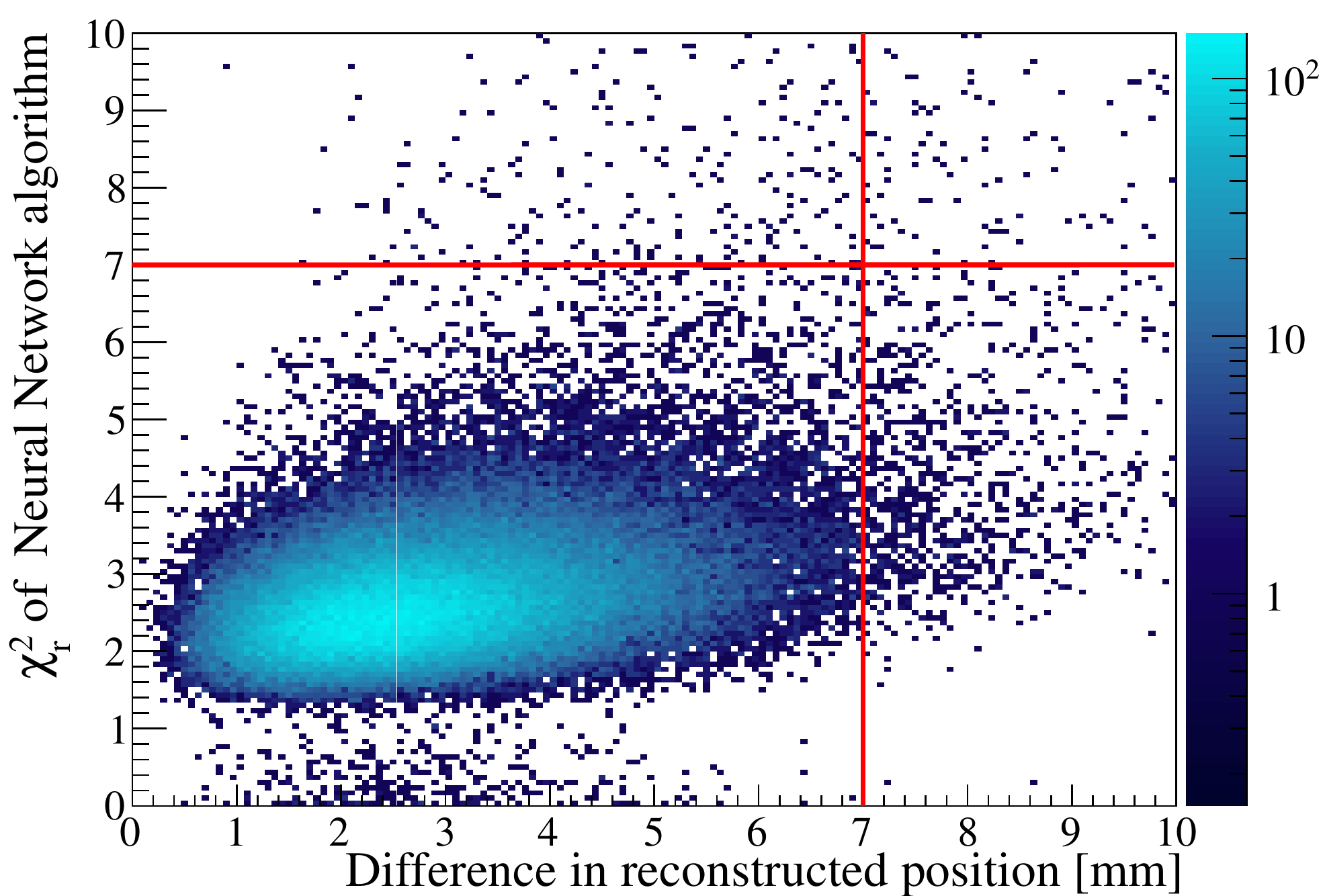}
\includegraphics[width=0.82\columnwidth]{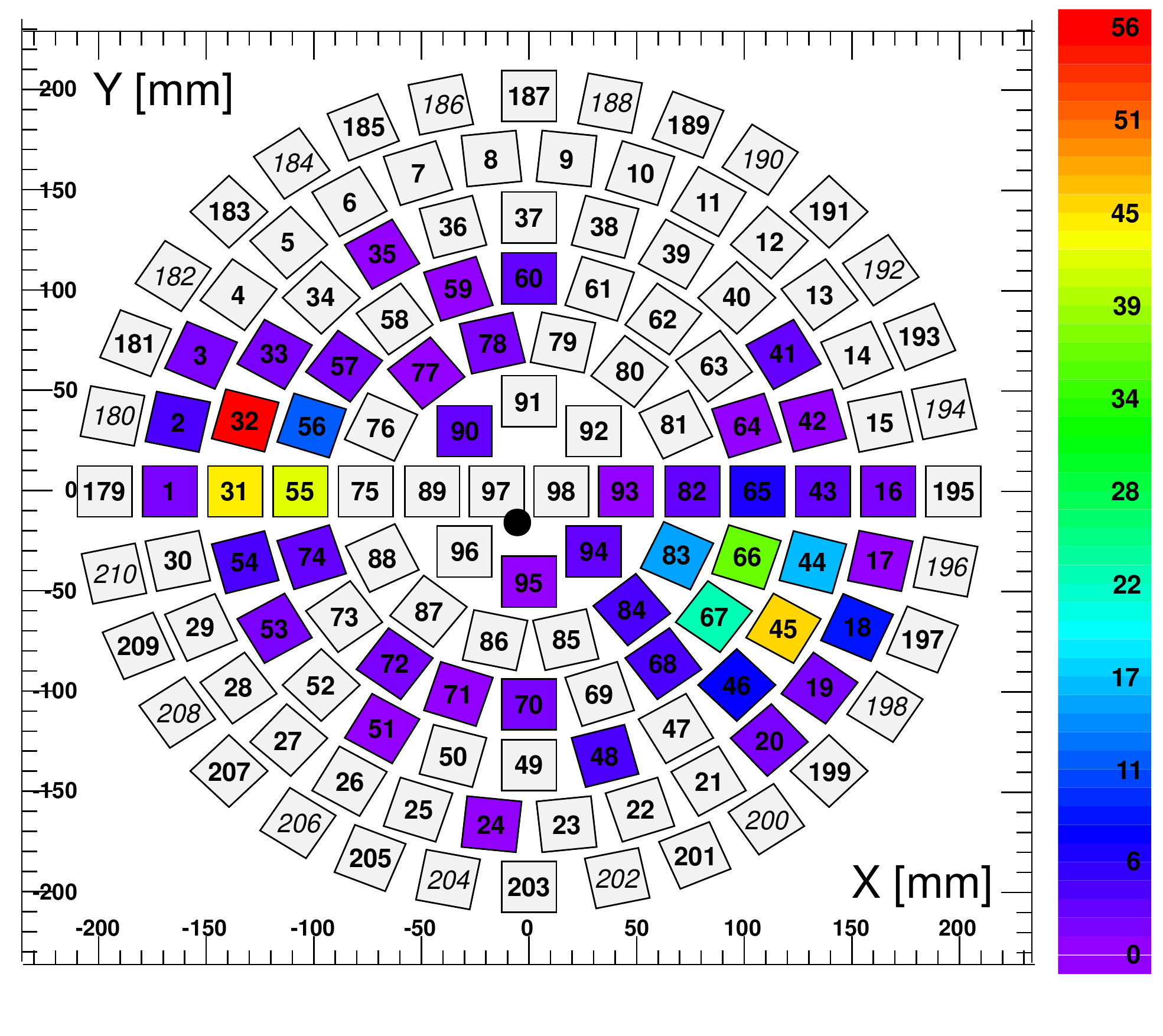}
\caption{{Top: Neural network $\chi^2_{r}$ value versus the combined difference in the reconstructed position by the NN, SVM and $\chi^2-$minimization reconstruction algorithm from the $^{241}$AmBe calibration data in Run-I. Bottom: example of a double scatter event from neutron data rejected by the position reconstruction cuts.  The PMT pattern of the top array clearly shows the presence of two S2 signals centered roughly on PMT 32 and 45, respectively. The black dot close to the center represents the reconstructed position. The color code of the legend represents the measured S2 signal size (in PE) seen by each PMT.}}
\label{fig:posrec}
\end{figure}

Large differences in the event positions inferred by the three different algorithms (see Sec.~\ref{EventRec}) in general correspond to unresolved multiple scatter events as their interactions are so close in $z$ that the corresponding S2 signals cannot be separated.  Events with the combined difference, calculated as the square-root of the sum of squared differences between $(x,y)$ positions reconstructed by NN, SVM and $\chi^2$-minimization algorithms, larger than 7\,mm are removed. A $\chi^2$-value of the difference between the observed S2 PMT hit pattern and the one expected from the Monte Carlo simulation quantifies the quality of the reconstruction position. Hence, a cut on the reduced $\chi^2_{r}$ value ($\chi^2$ divided by the number of PMTs in the upper array) of the  NN algorithm is used to reject badly reconstructed events. Figure~\ref{fig:posrec} (top) shows the $\chi^2_{r}$ distribution of events calculated from an NN algorithm versus the combined difference in reconstructed position. Events above and on the right of the solid lines,  are rejected. Figure~\ref{fig:posrec} (bottom) shows the S2 PMT pattern of a double scatter event which is rejected by these conditions. The position is reconstructed wrongly (black dot) close to PMT 97, in between the two true positions. {The acceptance of these conditions is evaluated using both ER and NR calibration data. It is $>$\,99.6\% in the energy ROI of both runs}.

\begin{figure}[!h]
\centering
\includegraphics[width=1\columnwidth]{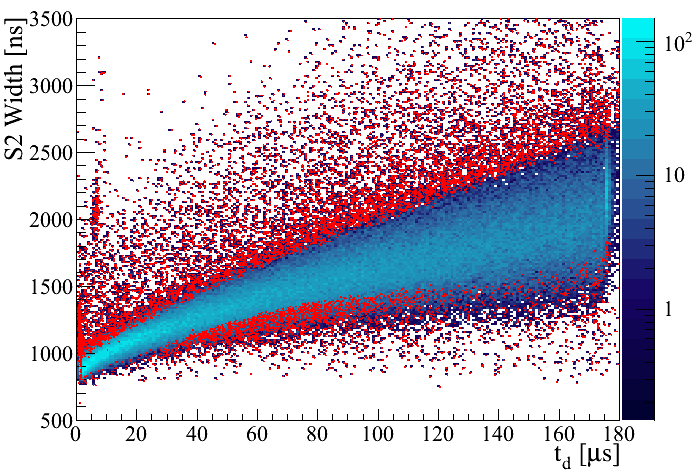}
\includegraphics[width=1\columnwidth]{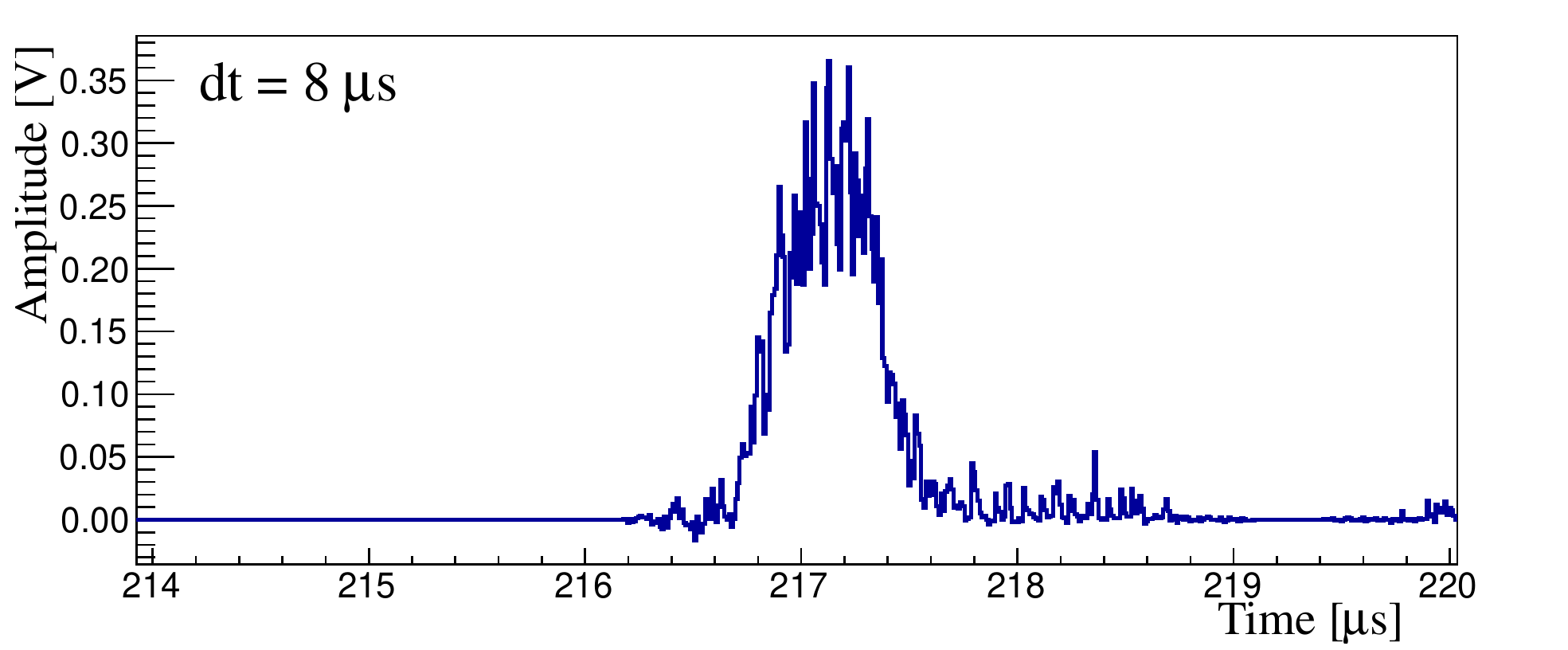}
\includegraphics[width=1\columnwidth]{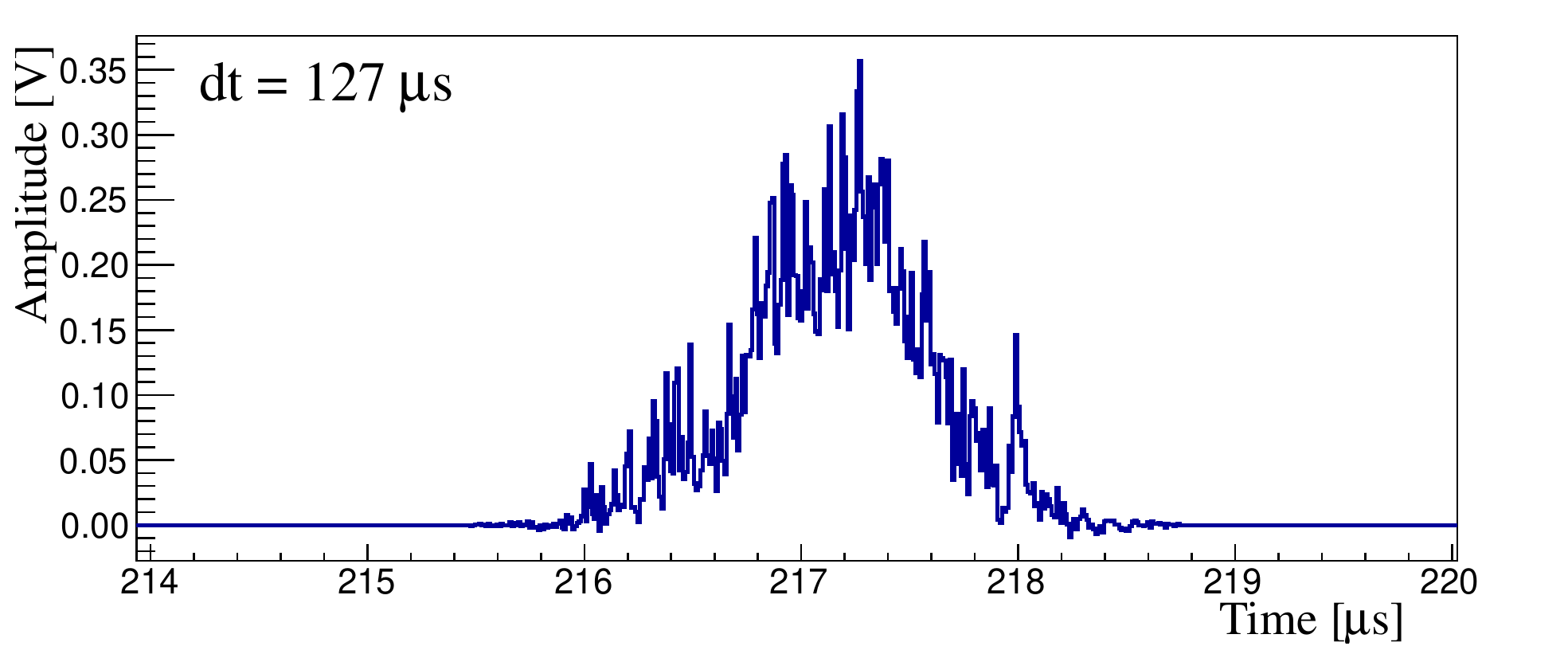}
\caption{Top: Distribution of the S2 width (defined at 10\,\% of the peak height) versus drift time for nuclear recoils with energies up to 30\,PE in S1 in Run-I (blue histogram). The red dots are the events removed by the S2 width selection.  Middle: S2 waveform example with $t_d =$\,8\,$\mu$s. Bottom: S2 waveform example with $t_d = $\,127\,$\mu$s.}
\label{fig:widthcut}
\end{figure}

The width of an S2 pulse increases with the $t_d$ due to the diffusion of the electron cloud during its propagation towards the gate grid. Typical width values, defined at 10\% of its peak height, range from $\sim1\,\mu$s to $\sim2\,\mu$s for minimal and maximal ($t_d = 176\,\mu$s) drift times, respectively.  The $t_d$-dependent S2 width is compared to the $t_d$ value and events are rejected when these quantities are inconsistent. Neutron calibration data are used for the cut definition. Low energy S2 events show larger spread due to low statistics of drifted ionization electrons, hence the cut is defined in an energy-dependent way. {There is a small $(x,y)$ dependence of the S2 width. This dependence was accounted for in Run-II, however no correction was done in Run-I.} The cut {definition is set to 90\% acceptance for Run-I and 92\% for Run-II (blue and red triangles in Fig.~\ref{fig:acc}, respectively) }. The top panel of Fig.~\ref{fig:widthcut} shows the distribution of the S2 width versus $t_d$ for nuclear recoil events from neutrons {up to 30\,PE (blue)}. The red dots are the events removed by the S2 width selection.  
The waveforms in the middle and bottom panels correspond to accepted typical candidates with $t_d =$\,8\,$\mu$s and $t_d =$\,127\,$\mu$s between the S1 and S2 peak. Their uncorrected S2 sizes are $\sim$1\,000\,PE and $\sim$1\,500\,PE, respectively.

{Multiple scatters from interactions, where one interaction happens inside the TPC and at least one in a charge insensitive region, can produce electronic or nuclear recoil events with an unusually low S2/S1 ratio. Such regions include the LXe volume below the cathode or close to the field shaping rings near the inner TPC wall. These events are classified as valid single scatters because a single S2 is measured but the S1 is the sum of both prompt S1 signals. Having an S2/S1 ratio lower than that of true ER/NR single scatters, the events can potentially leak into the WIMP search region (the so-called ``anomalous background'').} 

\begin{figure}[!h]
\centering
\includegraphics[width=1\columnwidth]{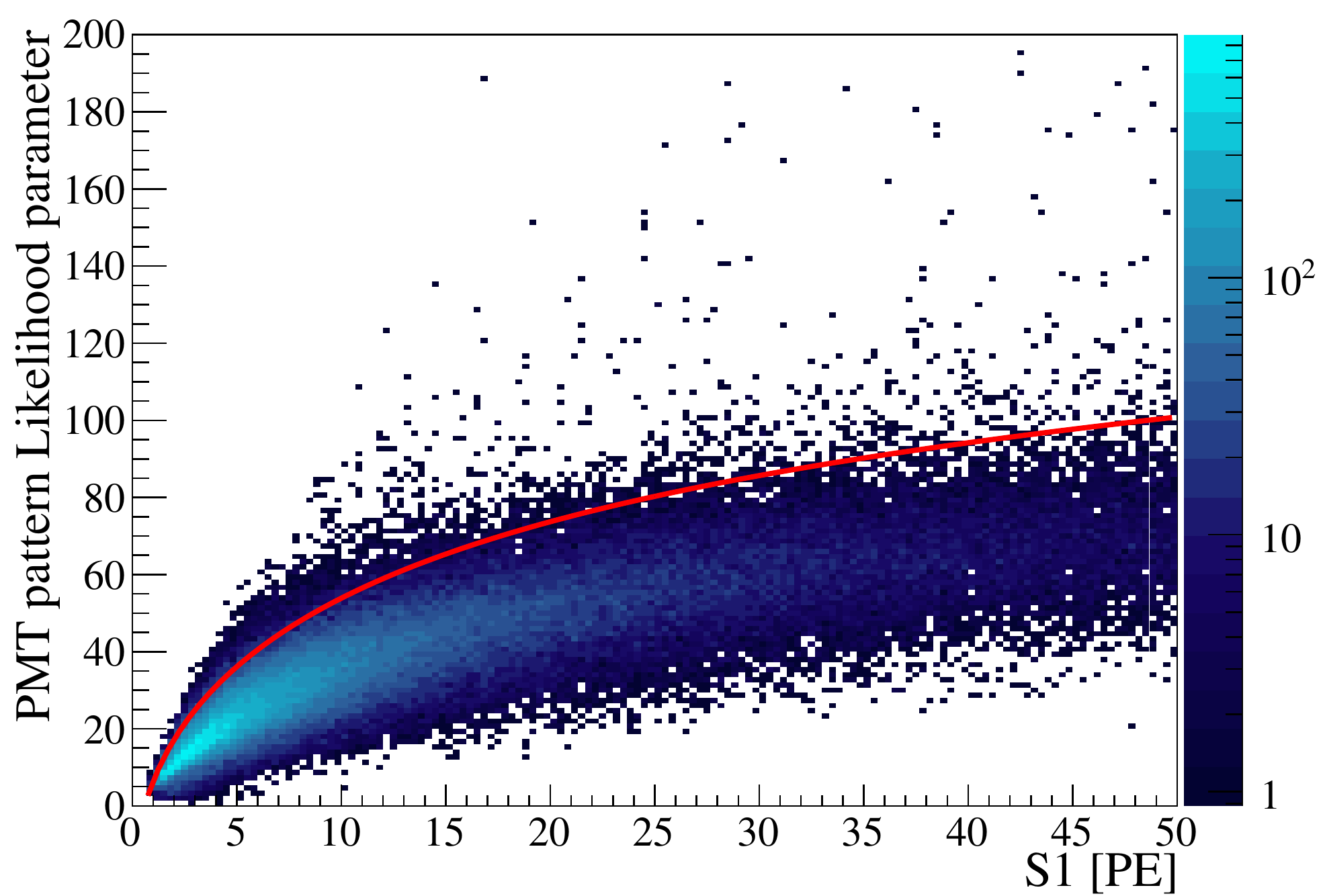}
\caption{Distribution of the likelihood parameter of the PMT pattern as a function of the uncorrected S1 signal for nuclear recoils in Run-I. The likelihood parameter is computed as $-2\log(\lambda_P)$, where $\lambda_P$ is the ratio of the Poisson likelihood of the measured S1 PMT pattern and the ``standard" S1 PMT pattern at the same position (see text). The red line represents the cut line above which events are discarded.}
\label{fig:anomalous}
\end{figure}

The PMT hit pattern of the S1 signal is used to discriminate between these events and true single interactions,  {using similar technique as described in\,\cite{Angle:IDM2009}.}
A likelihood parameter is computed as $-2\log(\lambda_P)$, where $\lambda_P$ is the ratio of the Poisson likelihood of the measured S1 PMT pattern and the ``standard" S1 PMT pattern of single scatter events which happen at the same reconstructed $(x,y,z)$ position. The standard S1 PMT pattern map is acquired from the full absorption peak of $^{137}$Cs calibration data. Figure~\ref{fig:anomalous} shows the distribution of the likelihood parameter for {Run-I} neutron data as a function of the uncorrected S1 signal area. Events with a poor likelihood ratio (above the line) are rejected because they have an S1 pattern that is inconsistent with the expectation based on their position. {The cut was defined on low-energy ER calibration data, the cut acceptance was determined with NR calibration data. The acceptance is $\sim$\,97\% with no energy dependence.}

\begin{figure}[h!]
\centering
\includegraphics[width=0.82\columnwidth]{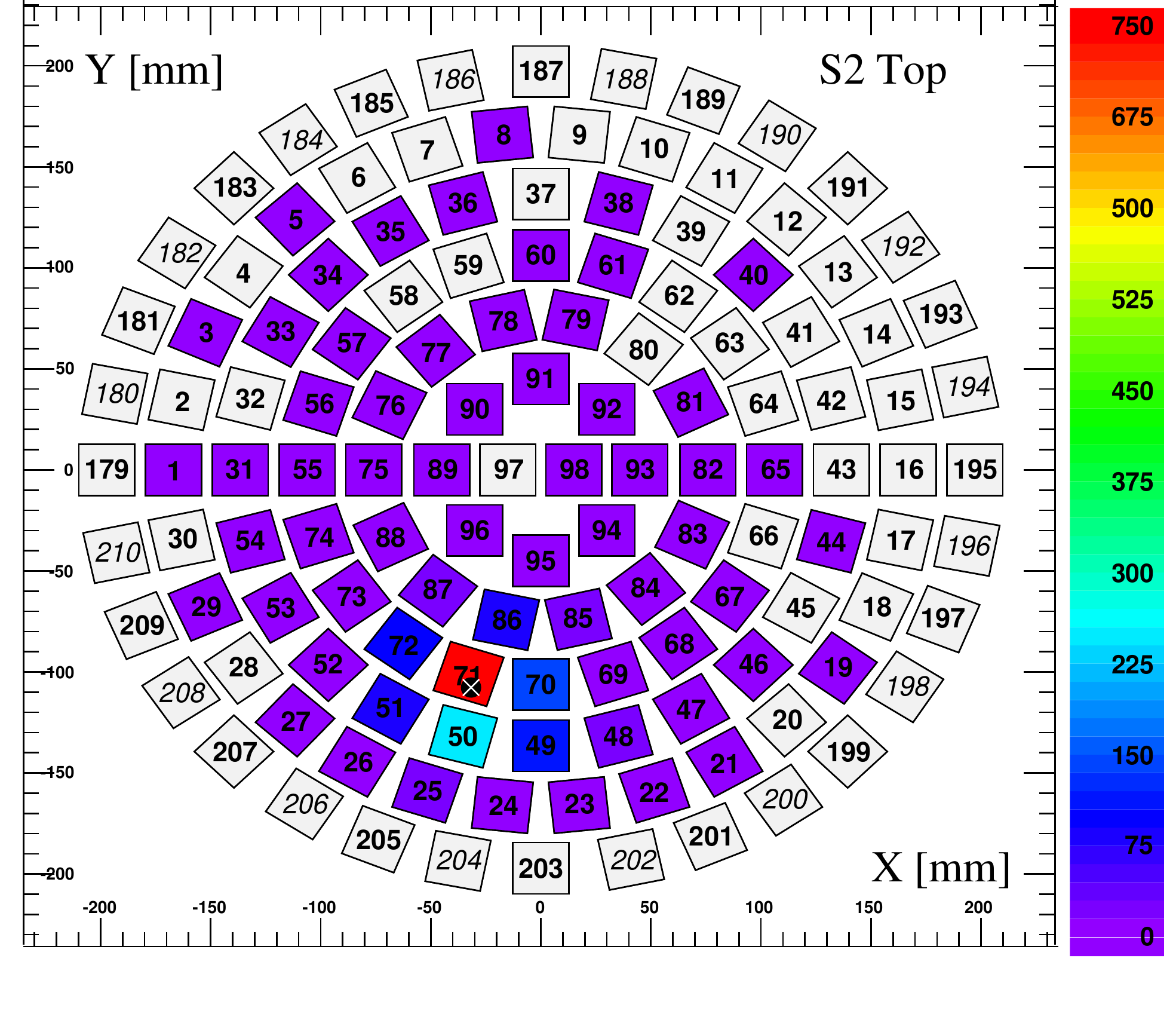}
\includegraphics[width=0.82\columnwidth]{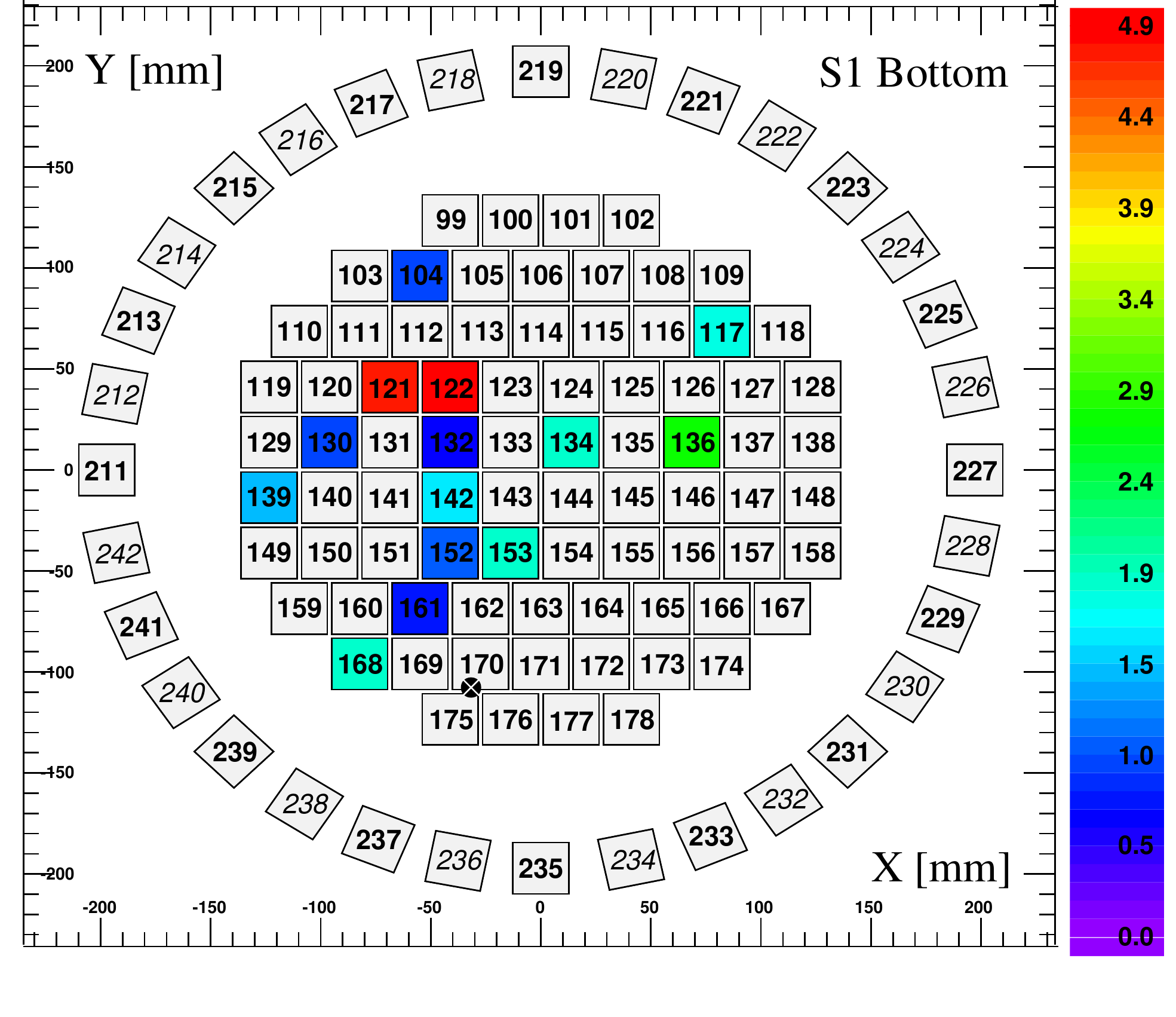}
\caption{{Example of a low-energy event (uncorrected S1\,=\,33\,PE) removed by the pattern likelihood selection (the event has $-2\log(\lambda_P)\,=\,96$, compare with Fig.~\ref{fig:anomalous}). Top: The top array PMT pattern of the S2 signal with the reconstructed ($x,y$) position (black dot). Bottom: The bottom array PMT pattern of the S1 signal.  This event could be due to a double scatter. One interaction happens below the cathode and close to the bottom PMT array resulting in a very localized S1 above PMTs 121 and 122 with no corresponding S2 signal at similar ($x,y$). The second interaction happens in the sensitive volume 8.53\,cm above the cathode. Its ($x,y$) position is well reconstructed using the S2 signal in the top array.}}
\label{fig:anomalousPat}
\end{figure}

{Figure~\ref{fig:anomalousPat} shows an example of a low-energy event removed by the pattern likelihood selection which could be explained by a double scatter event with one interaction  below the cathode. The top panel contains the pattern of the S2 signal in the top PMT array. The ($x,y$) position is well reconstructed close to PMT 71 (black dot). The bottom panel shows the pattern of the S1 signal in the bottom PMT array. There is an unusually large portion of the S1 signal seen right above PMTs 121 and 122 pointing to a possible second scatter in the vicinity of the cathode.}

\subsubsection{Signal/background discrimination and fiducial volume}\label{DiscrCus}% - - - - -  -

{This section uses Run-I as an example of how the signal/background discrimination parameters and fiducial target mass were established. A similar analysis was performed for Run-II, see~\cite{Aprile:2012kx}.} 

\begin{figure}[!h]
\centering
\includegraphics[width=1.0\columnwidth]{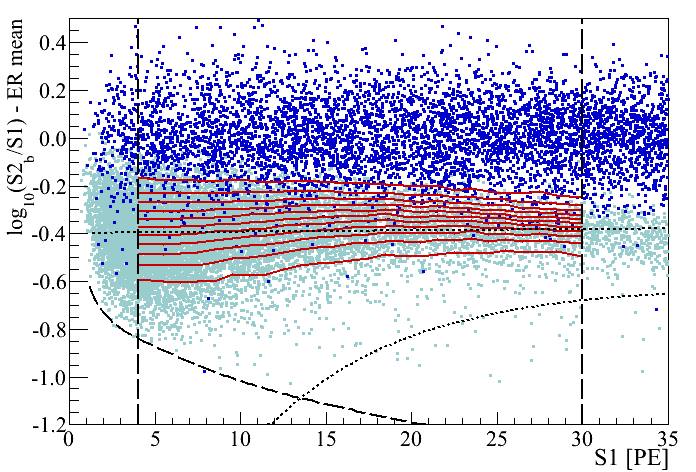}
\caption{Electronic (blue) and nuclear (cyan) recoil bands in flattened discrimination space {for Run-I}. 
The black dashed lines, representing the analysis energy window (vertical) and the S2 threshold cut (bottom left), are used in both analysis methods. The 12 bands used in the Profile Likelihood analysis to model signal and background are shown in red. The benchmark WIMP region for the cut-based analysis is further constrained by the black dotted lines  which are 99.75\% electronic recoil rejection line (middle) and  the lower 3\,$\sigma$ contour of the neutron distribution (bottom right).}
\label{fig:bands}
\end{figure}

Figure~\ref{fig:bands} shows the electronic and nuclear recoil bands from $^{60}$Co (blue)  and $^{241}$AmBe (cyan) data, respectively, using the discrimination parameter $\log_{10}(\n{S2_b/S1})$, flattened by subtracting the electronic recoil mean. This removes the energy dependence of the electronic recoil band allowing for an easier combination of data from different energies. The border lines of the 12 bands used by the Profile Likelihood analysis to model signal and background distributions are shown in red in the chosen energy window. 
 
Compared to the Profile Likelihood analysis, the benchmark WIMP region for the cut-based analysis is additionally constrained by the S2/S1 electronic recoil rejection cut and the lower bound of the nuclear recoil band. The electronic recoil rejection level was defined using $^{60}$Co calibration data which {have} an S2/S1 distribution that is described well by a Gaussian function with an extra tail appearing about 3\,$\sigma$ away from the mean value. It was set to 99.75\% (see Fig.~\ref{fig:bands}). The acceptance of this cut, calculated using nuclear recoils, is energy dependent and ranges from $\sim$\,29\% to $\sim$\,46\% (see Fig. 2 of~\cite{Aprile:2011hi}). The $^{241}$AmBe data {are} used to constrain the signal region by removing events which are more than 3\,$\sigma$ away from the mean of the nuclear recoil band (see Fig.~\ref{fig:bands}).

\begin{figure}[!h] 
\centering
\includegraphics[width=1.0\columnwidth]{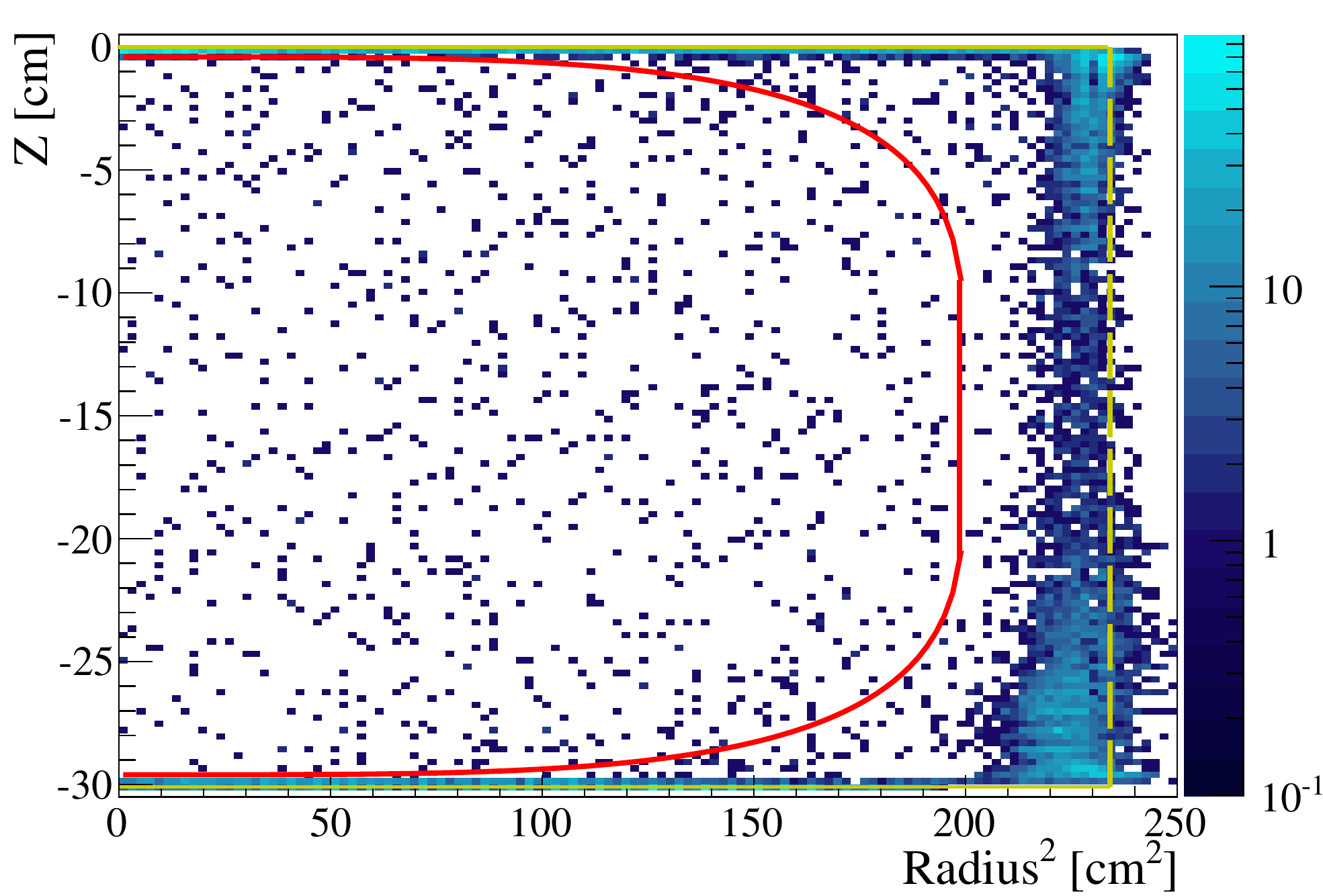}
\caption{Distribution of all events  in the TPC observed in the $8.4-44.6$\,keV${_{\n{nr}}}$ energy range during {Run-I}. With the exception of the electronic recoil discrimination cut, all cuts are used  including the ones introduced post-unblinding to remove a population due to electronic noise.  The fiducial volume containing 48\,kg of liquid xenon (red) and the TPC borders (yellow) are also indicated. Events outside of the borders are due to the uncertainty in the position reconstruction.} \label{fig:fv_result}
\end{figure}

An ellipsoid containing 48\,kg of liquid xenon is used as the fiducial target mass {in Run-I}. Figure~\ref{fig:fv_result} shows the shape of the fiducial volume cut (red) together with the observed event distribution from the science data without applying any electronic recoil discrimination cut.  The yellow dashed-contour {represents} the borders of the TPC. Events falling outside of it are due to the uncertainty in the position reconstruction.
The background during {Run-I} was dominated by $^{85}$Kr which is uniformly distributed in the TPC as is visible in the central region of Fig.~\ref{fig:fv_result} (see also Sec.~\ref{ERbackground}). For this reason, a smaller fiducial volume would not help to reduce the background further, and the largest possible volume was used instead. Only the edges of the detector were excluded as signal corrections and position reconstruction uncertainties are largest here. In the remaining region the volume was enlarged until the background from the detector components increased significantly.
The $\log_{10}(\n{S2_b/S1})$ distribution shows a population of events at low energies (few PE) and high values of the discrimination parameter, $\log_{10}(\n{S2_b/S1}) - $\,ER mean\,$>$\,0.5. These correspond to random small S1 peaks that do not belong to the S2 signal. A dedicated cut removes these events which are far above the nuclear recoil band with an acceptance above 99.95\%.

{The uniformly distributed $^{85}$Kr background in Run-II was much lower than in Run-I (see Sec.~\ref{ERbackground}) and this allowed to improve the signal-to-background by reducing the fiducial mass to 34\,kg.}

\subsubsection{Combined cut acceptance} % ===================================

\begin{figure}[h]
\centering
\includegraphics[width=1\columnwidth]{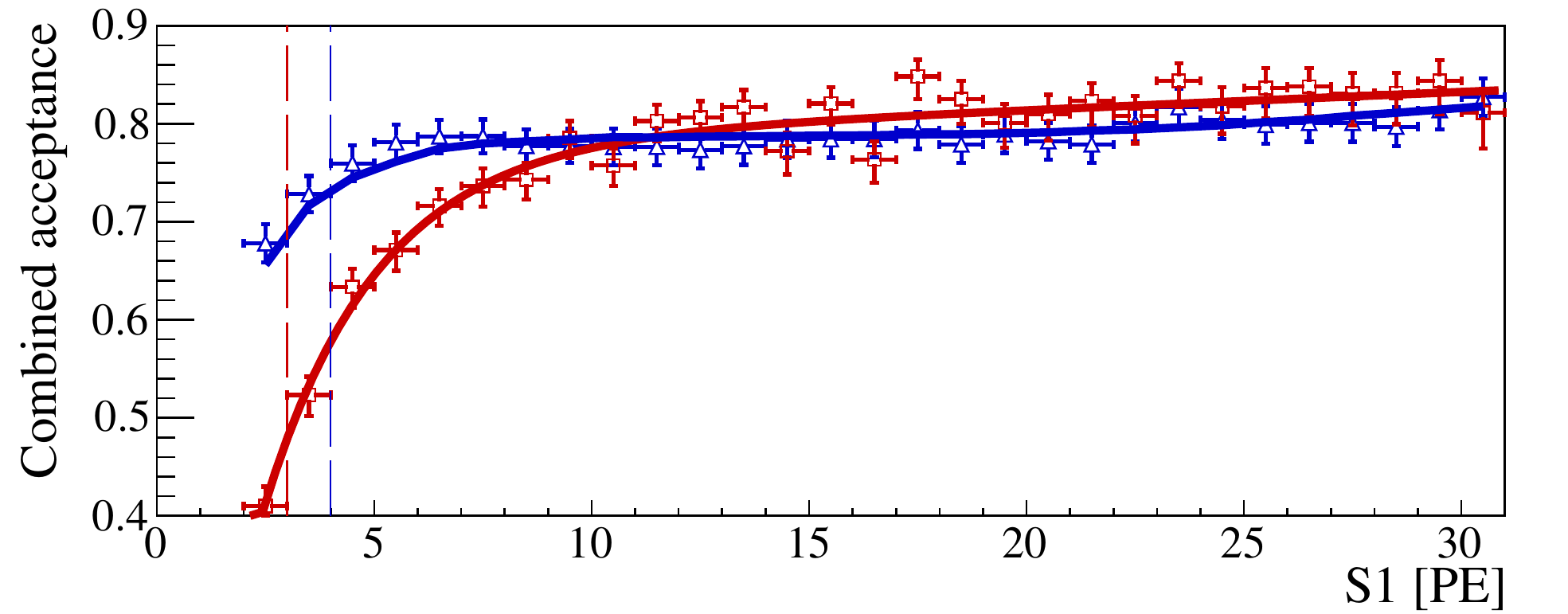}
\caption{{Combined acceptance of the quality cuts as a function of the corrected S1 signal, for Run-I (blue triangles) and Run-II (red squares), including the smoothed curves used in the PL analysis (lines). The difference at low S1 is explained by the different methods to derive the acceptance of the S1 coincidence requirement, using a Monte Carlo method (Run-I) and a superior data-driven method (Run-II), see Fig.\,\ref{fig:acc}. The S2 threshold cut is treated separately (see Fig.~\ref{fig:accS2thr}). The vertical dashed lines correspond to the lower energy threshold for Run-I (blue) and Run-II (red).}}
\label{fig:acc2}
\end{figure}

The cumulative acceptance of all cuts used for the Profile Likelihood analyses in the {S1 ROI} energy range and {science-run-dependent} fiducial mass is shown in Fig.~\ref{fig:acc2}. {The blue data points represent Run-I with the S1 coincidence requirement derived by a Monte Carlo simulation. The red points show the combined acceptance of Run-II using the improved data-driven method, see Fig.~\ref{fig:acc}. The acceptance of the S2 threshold is shown in Fig.~\ref{fig:accS2thr}. The latter has to be applied first to the WIMP spectrum without statistical fluctuations.} Then the combination of all other cuts as shown in Fig.~\ref{fig:acc2} is applied to the S1 spectrum after taking into account Poisson fluctuations. As already mentioned, for the cut-based analysis, the S2/S1 electronic recoil rejection cut and the lower nuclear contour are applied additionally (see Fig.\,2 of~\cite{Aprile:2011hi} and Fig.\,1 of \cite{Aprile:2012kx}).

\subsection{Background expectation} \label{Background}

This section describes the various background contributions and quantifies the number of expected background events for both science runs, for the used fiducial target mass, exposure, cuts, and WIMP search benchmark region (see Table~\ref{tab:AnalysisPars}). In both Run-I and Run-II, the Profile Likelihood analyses used to derive the final results employ exactly the same data and assumptions for the background estimation.

%%%%%%%%%%%%%%%%%%%%%%%%%%%%%%%%%%%%%%%%%%%%%%%%%%%%%%%%%%%%%%%%%
\subsubsection{Nuclear recoil background from neutrons}

The nuclear recoil background from muon-induced neutrons and from neutrons created in spontaneous fission and ($\alpha$,n) reactions in the detector materials are determined using a GEANT4-based\,\cite{geant4:2003} Monte Carlo simulation\,\cite{Aprile:2013tov}. Single scatter nuclear recoils in the LXe  with no associated electronic recoil from inelastic reactions were selected assuming 3\,mm event resolution. 

Cosmogenic muons were simulated using the MUSIC\,\cite{antonioli-1997} and 
MUSUN\,\cite{Kudryavtsev09} packages. The muons and their daughter particles from the 
electromagnetic and hadronic showers were propagated through the rock and through 
the XENON100 shield and detector materials. A total of ($0.08^{+0.08}_{-0.04}$) and 
($0.12^{+0.12}_{-0.07}$)\,muon-induced events were expected for Run-I and Run-II, respectively.

To obtain the neutron rate from natural radioactivity, the $\alpha$-activity from the 
uranium and thorium chains is considered. These are derived from the measured detector material's 
$\gamma$-activities\,\cite{Aprile:2011ru}. The neutron production rates in the detector and shield 
materials, as well as the energy spectra from ($\alpha$,n) reactions and spontaneous fission, are 
calculated using a modified SOURCES4A code\,\cite{wWilson,Carson:2004cb}. For Run-I we expect a radiogenic neutron background of ($0.032\pm0.006$)\,events 
and ($0.052\pm0.010$)\,events for Run-II.

Summing up the two contributions, the expected total nuclear recoil backgrounds are 
($0.11^{+0.08}_{-0.04}$) and ($0.17^{+0.12}_{-0.07}$)\,events for Run-I and Run-II, respectively,
dominated by cosmogenic neutrons in both cases. The rather large 
errors are due to systematic uncertainties in the neutron production rates\,\cite{Aprile:2013tov}.

%%%%%%%%%%%%%%%%%%%%%%%%%%%%%%%%%%%%%%%%%%%%%%%%%%%%%%%%%%%%%%%%%
\subsubsection{Electronic recoil background from $\gamma$ and $\beta$ events} \label{ERbackground}

\begin{figure}[h]
\centering
\includegraphics[width=1\columnwidth]{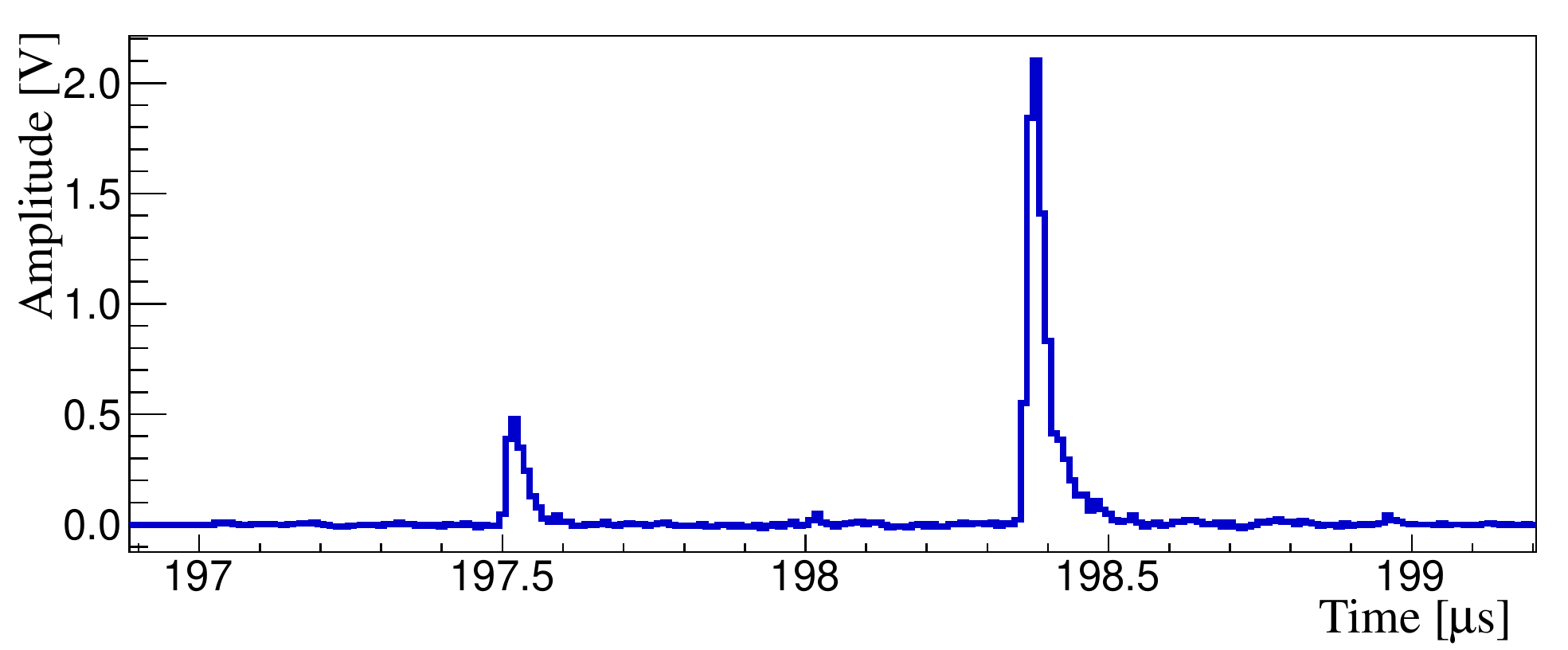}
\caption{S1 peaks of a candidate $^{85}$Kr event where the second light signal 
from the $\gamma$-ray is delayed by $\sim 900$\,ns.}
\label{fig:kr85}
\end{figure}

The electronic recoil background conditions were different in the two science runs considered 
here, leading to slightly different ways to predict the background before unblinding the data.
During the 100.9\,live days of Run-I, the background at low energies was dominated 
by intrinsic $^{85}$Kr ($\beta$ with $E_0=687$\,keV$_{\n{ee}}$ endpoint energy), introduced 
accidentally through a tiny air leak in the gas system before the start of the dark matter search.  
The $^{85}$Kr concentration was determined by comparing the measured energy spectrum to a 
detailed Monte Carlo simulation. The simulation includes the intrinsic radioactivity of 
all detector components, based on the measured contamination of the various materials. 
This method is described in detail in~\cite{Aprile:2011vb}. 
Assuming a $^{85}$Kr isotopic abundance of $2 \times 10^{-11}$\,\cite{Du:2004}, a natural Kr~concentration of $(350 \pm 50)$\,ppt was obtained. In parallel, an independent analysis using delayed coincidences was performed\,\cite{Alimonti:2002xc}. It uses the decay of $^{85}$Kr to $^{85m}$Rb ($\beta$ with $E_0=173.4$\,keV$_{\n{ee}}$) followed by the decay of $^{85m}$Rb to $^{85}$Rb ($\gamma$ of 514\,keV$_{\n{ee}}$) with 1.46\,$\mu$s half-life and a branching ratio of 0.434\%. Figure~\ref{fig:kr85} shows the waveform of a $^{85}$Kr candidate event. The natural Kr concentration of ($294 \pm 66$)\,ppt, derived with this method, is compatible with the one inferred from the data/Monte Carlo comparison. A direct measurement using rare gas mass spectrometry (RGMS) yields a moderately higher value of ($450 \pm 30$) ppt.

In Run-II, the $^\n{nat}$Kr concentration was lowered by cryogenic distillation. The values measured via RGMS and delayed coincidences were ($19\pm4$)\,ppt and ($18\pm8$)\,ppt, respectively. Because the concentration is now subdominant compared to the intrinsic contamination with Rn, which was measured to be 65\,$\mu$Bq/kg by tagging high-energy $\alpha$-signals, it cannot be obtained anymore from spectral fits of the background.

The electronic recoil background in the signal region consists of two contributions: leakage from the Gaussian shaped bulk of the electronic recoil background distribution and a small contribution of anomalous (non-Gaussian) leakage. Since intrinsic $\beta$-decays from $^{85}$Kr are dominating the background of Run-I while being negligible in Run-II, the way to predict the expected number of background events in the WIMP search region is slightly different.  

In Run-I, the Gaussian component is determined by extrapolating the low-energy science data of the non-blinded  $\log_{10}(\n{S2_b/S1})$ region. The prediction is (1.14$\pm$0.48)\,events with the error being mainly due to the uncertainty in the definition of the electronic recoil discrimination level.

Events with incomplete charge collection as defined in Sec.~\ref{DiscrCus} contribute to the anomalous background which cannot be described by a Gaussian distribution. Spatial and spectral distribution of anomalous leakage events in the science and $^{60}$Co calibration data may not be identical because they originate from different sources located at different positions. Hence it was studied whether $^{60}$Co data could be used to predict the anomalous background. Anomalous events with one scatter in a charge insensitive region and a subsequent low S2/S1 value are also expected to populate the S2/S1 region above the blinding cut. These events were selected using the S1 pattern likelihood parameter in the science and $^{60}$Co data. Their spectral and spatial distributions agree at the 10\% level. In addition, the spatial distributions of events with incomplete charge collection in Monte Carlo simulations of background and $^{60}$Co calibration runs were compared and also found to be similar within 10\% in the energy region of interest and the selected fiducial volume. These remaining discrepancies are taken into account in the systematic uncertainty of the anomalous background. 

$\beta$-decays from the dominating $^{85}$Kr cannot create spatially-resolved multiple scatters due to the short electron mean free path, which is much less than the 3\,mm position resolution. Taking into account the different exposure of the $^{60}$Co calibration compared to the science data, and the fraction of the background due to $\gamma$-interactions that can undergo multiple scatters, the anomalous leakage into the benchmark region was estimated to be ($0.56^{+0.21}_{-0.27}$)\,events. The uncertainty in the $^{85}$Kr concentration is also considered in the error.

In Run-II, much more ER calibration data was acquired using $^{60}$Co and $^{232}$Th sources. No difference was observed in the response of the two sources, also in the tails, and therefore both were used for the prediction of the background, which was no longer $^{85}$Kr-dominated. The expected number of events from ER background was determined in the following way. Calibration data were scaled to the dark matter data by normalizing to the number of events above the blinding cut and by counting the number of calibration events remaining in the benchmark region. This leads to the prediction of ($0.79 \pm 0.16$)\,events. For the background model used in the Profile Likelihood analysis, the same calibration data were separated into a Gaussian and an anomalous part. The latter was obtained by subtracting the Gaussian contribution and by describing the the remaining events by a 2-dimensional function which was constant along the discrimination-axis $\log_{10}$(S2$_b$/S1)$-\n{ER}_\n{mean}$ and exponential along the S1-axis.\\[0.4cm]

%%%%%%%%%%%%%%%%%%%%%%%%%%%%%%%%%%%%%%%%%%%%%%%%%%%%%%%%%%%%%%%%%
\subsubsection{Total background expectation and side-band test}

Taking into account all background sources mentioned above, the total background predictions 
for the benchmark regions of Run-I (100.9\,days $\times$ 48\,kg) and Run-II 
(224.6\,days $\times$ 34\,kg) are $(1.8\pm0.6)$\,events and $(1.0\pm0.2)$\,events, respectively. 
Table~\ref{tab:BGcomp} summarizes the background contributions.
$\alpha$-decays in the LXe do not contribute to the background as they have energies of a 
few MeV, far above the energy region of interest for the dark matter search.

\begin{table}[h]
\caption{Contributions to the background prediction for the benchmark WIMP search 
regions in both science runs, taking into account the correct exposure and acceptance.}
\label{tab:BGcomp}
\begin{tabular}{lcc}
\hline
\hline
\multirow{2}{*}{Source} & \multicolumn{2}{c}{Expected background [events]} \\
& Run-I & Run-II \\
\hline
ER Gaussian leakage& 1.14$\pm$0.48 & \multirow{2}{*}{0.79$\pm$0.16} \\
ER anomalous leakage& $0.56^{+0.21}_{-0.27}$ & \\
NR from neutrons & $0.11^{+0.08}_{-0.04}$ & $0.17^{+0.12}_{-0.07}$ \\
\hline
Total & 1.8$\pm$0.6 & 1.0$\pm$0.2 \\
\hline
\hline\\
\end{tabular}
\end{table}

Before analyzing the science data of Run-I in the region of interest, the side-band  region ($30-130$)\,PE was unblinded. The background predictions for the 99.5\%, 99.75\% and 99.9\% electronic recoil discrimination levels of Run-I were $14.3^{+7.7}_{-8.5}$, $8.2^{+2.2}_{-2.6}$, and $5.6^{+0.6}_{-1.0}$~events while 10, 8, and 7~events were observed, respectively. The observed and predicted numbers of events are compatible within the errors. For Run-II, the ($30-100$)\,PE region was used and $(6.0 \pm 0.4)$, $(3.2 \pm 0.3)$, $(1.6 \pm 0.2)$ events were expected for the three discrimination levels, while 4, 2, and 1 events were observed. Also in this case expectation and observation are in statistical agreement, confirming the validity of the background predictions before unblinding the WIMP search region.

\subsection{Observed event population}\label{EventPopulation}

{This section provides an overview of the observed event population, using Run-I as an example. The observed event population for Run-II can be found in~\cite{Aprile:2012kx}.}

\begin{figure}[h]
\centering
\includegraphics[width=1.0\columnwidth]{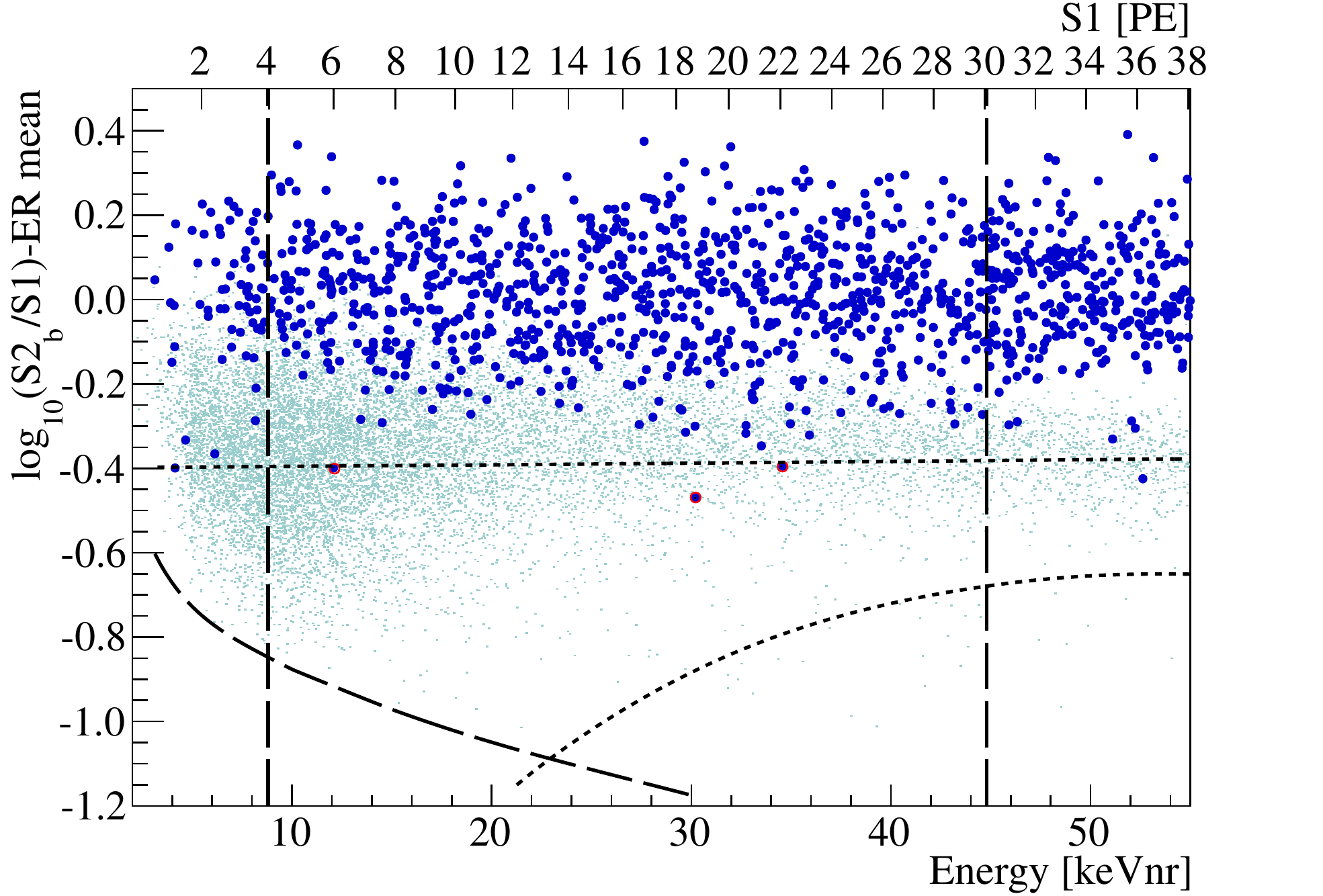}
\caption{The electronic/nuclear recoil discrimination parameter  as a function of nuclear recoil energy {for Run-I}. 
Blue points indicate the observed event distribution after all cuts.  Cyan points show the nuclear recoil distribution as measured with an $^{241}$AmBe neutron source. The dashed-lines represent the energy window $8.4-44.6\1{keV_{\n{nr}}}$ and the threshold cut $\n{S2}>300\1{PE}$. The dotted-lines indicate the 99.75\% rejection line from above  and the $3\sigma$ contour of the NR distribution from below. These are not used in the Profile Likelihood analysis, but are used to define the benchmark WIMP region. Three events with energies
of 12.1, 30.2 and 34.6\,$\1{keV_{\n{nr}}}$ fall into the benchmark region (red circles). This plot shows the same event distribution as Fig.\,3 in\,\cite{Aprile:2011hi}.}\label{fig:results}
\end{figure} 

\begin{figure}[!h]
\centering
\includegraphics[width=1.0\columnwidth]{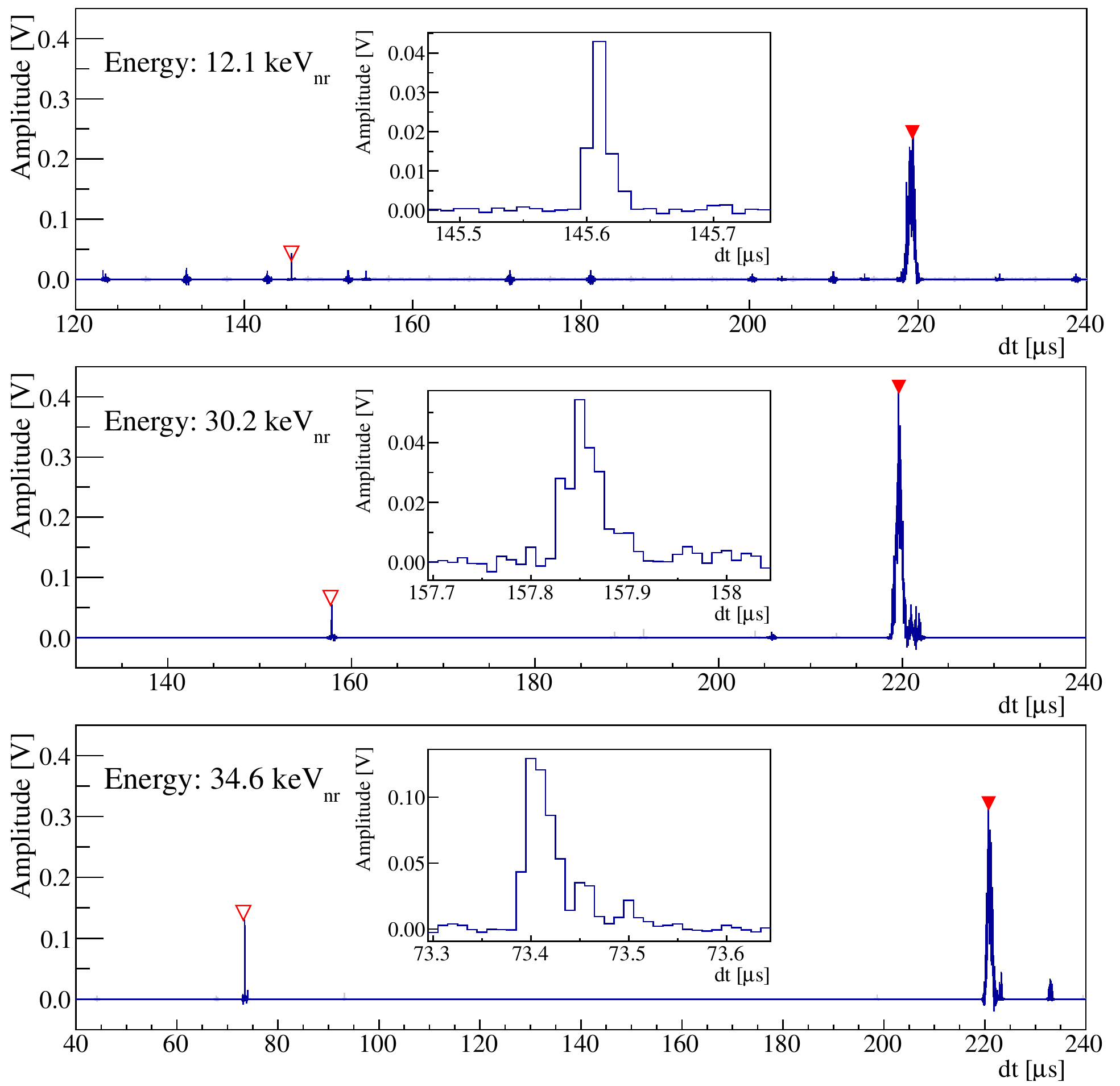}
\caption{Waveforms of the three {Run-I} events that pass all cuts in the benchmark region, including the ones defined post-unblinding. The S1 peaks are labelled  by open triangles, while the S2 peaks are labelled by full triangles. The insets show an expanded view around the S1 region. Periodic electronic noise is visible in the top event, while single electron pulses after the major S2 at the trigger position (220\,$\mu$s) are visible in the bottom event.}
\label{fig:GoodEvents}
\end{figure}

\begin{figure}[!h]
\centering
\includegraphics[width=1.0\columnwidth]{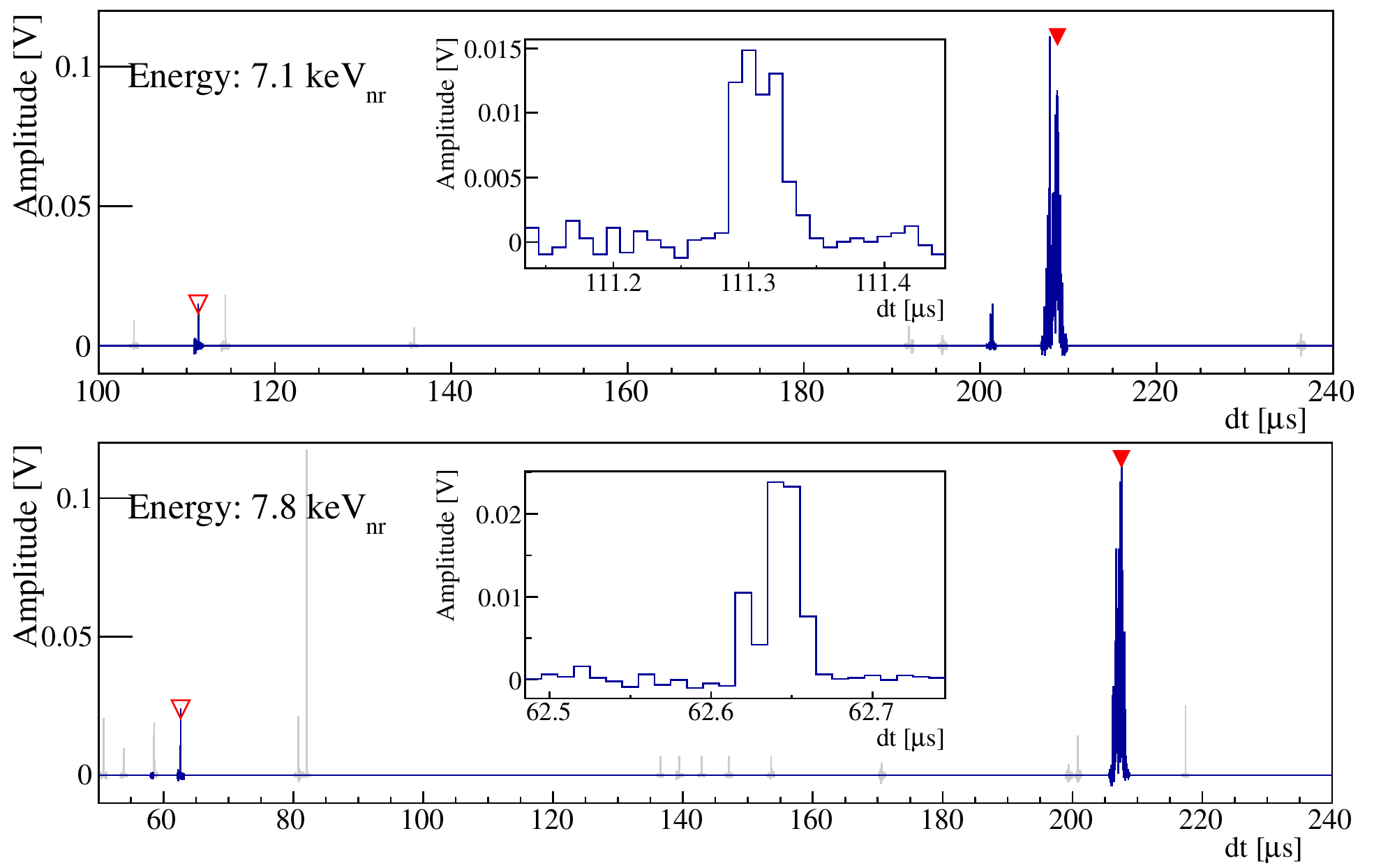}
\caption{{Waveforms of the two Run-II events that pass all cuts in the benchmark region. See Fig.~\ref{fig:GoodEvents} for details.}}
\label{fig:GoodEventsRun10}
\end{figure}

Figure~\ref{fig:results} shows the final distribution of events in the electronic/nuclear recoil discrimination parameter {for Run-I} (see Sec.~\ref{DiscrCus}) as a function of nuclear recoil energy $E_{nr}$. Based on the comparison of the observed event distribution  with the background-like distribution from Monte Carlo simulations, the background-only hypothesis was tested using a Profile Likelihood method\,\cite{Aprile:2011hx}. The $p$-value for this hypothesis, which is the probability that the outcome of a hypothetical  XENON100 experiment  results  in an observation equal or less background-like than the observed one, was found to be 31\%.

In the pre-defined benchmark WIMP search region, three events remain after all cuts. These events with energies of 12.1, 30.2 and 34.6\,$\1{keV_{\n{nr}}}$, as derived from their S1 signal, are marked by red circles in Fig.~\ref{fig:results}. The events are homogeneously distributed inside the fiducial volume, as can be seen in Fig.\,4 of reference\,\cite{Aprile:2011hi}. Their waveforms are shown in Fig.~\ref{fig:GoodEvents}. 
Though one of the events shows presence of periodic electronic noise in the waveform (top panel), the S1 signal has a higher amplitude and is not affected by it. The other two events have no noise present in their waveform (middle and bottom panels).  

{Within the background expectation of $(1.8\pm0.6)$ events in the cut-based analysis, the 3 observed events in Run-I do not constitute evidence for dark matter as the probability of a Poisson process with an expectation value of 1.8~events to fluctuate up to 3 or more events is 28\%. A similar analysis performed in Run-II yielded a background expectation of $(1.0\pm0.2)$ events for the cut-based analysis. Two events were observed in the signal region after unblinding, see Fig.~\ref{fig:GoodEventsRun10} for their waveforms. Here the Poisson probability was 26\%; in both runs the Poisson probability is consistent with the $p$-value of the Profile Likelihood model which gives no significant excess of events over the expected background. Using the main Profile Likelihood analysis, exclusion limits on spin-independent  elastic and inelastic WIMP-nucleon scattering interaction were placed in \,\cite{Aprile:2011hi} and \,\cite{Aprile:2011ts} for Run-I. Similarly, limits on spin-independent and spin-dependent elastic WIMP-nucleon scattering interactions were placed in \cite{Aprile:2012kx} and \cite{Aprile:2013doa} for Run-II.}

\section{Summary and Outlook} % === = = = = = ====== = = = = = ====== =

A detailed description of the analysis of {two science runs acquired by XENON100 in 2010--2012}  is presented. The data set, the selection criteria, the evaluation of the cut acceptances and the background estimates 
have been described. {These data have} been used to derive results in terms of spin-independent elastic \,{\cite{Aprile:2011hi,Aprile:2012kx}, spin-dependent elastic~\cite{Aprile:2013doa}} and inelastic\,\cite{Aprile:2011ts} WIMP-nucleon interactions. 
{The} methods presented here are general in nature and also apply to {future} XENON100 dark matter search data. A further refinement of the analysis, an improved detector performance{, additional low energy calibration sources} and a greater exposure will possibly increase the sensitivity of the XENON100 experiment.

% =============================================================

\section{Acknowledgements}

We gratefully acknowledge support from NSF, DOE, SNF, Volkswagen Foundation, FCT, R\'egion des Pays de la Loire, STCSM, NSFC, DFG, MPG, Stichting voor Fundamenteel Onderzoek der Materie (FOM), the Weizmann Institute of Science and the EMG research center. T.M.U. acknowledges the Alexander von Humboldt Foundation for support through a Feodor Lynen scholarship. We are grateful to LNGS for hosting and supporting XENON100.

\end{document}